\documentclass[twocolumn]{aastex63}

\usepackage{amssymb}
\usepackage{amsmath}
\usepackage{amsfonts}
\usepackage{graphicx}
\usepackage{color}
\usepackage{hyperref}
\usepackage[caption=false]{subfig}

\def\lsim{~\rlap{$<$}{\lower 1.0ex\hbox{$\sim$}}}
\def\bsim{~\rlap{$>$}{\lower 1.0ex\hbox{$\sim$}}}

\def\hkmsmpc{\ {\rm km\,s^{-1}\,{\it h}Mpc^{-1}}}

\def\hmpc{\ {\rm {\it h}^{-1}Mpc}}

\def\hmsun{\ {\rm {\it h}}^{-1}M_\odot}

\def\hmmpc{\ {\rm {\it h}Mpc^{-1}}}

\def\Kel{\ {\rm K}}

\def\ergss{\ {\rm erg\, s^{-1}}}

\def\apjl{Astrophys.\ J.\ Lett.}

\def\aap{Astron.\ Astrophys.}

\def\apjs{Astrophys.\ J. Supp.}

\newcommand{\Halpha}{\mathrm{H}\rm{\alpha}}

\allowdisplaybreaks

\usepackage[normalem]{ulem}

\usepackage{lineno} 

\begin{document}
\title{Correlation between $\Halpha$ emitters and their cosmic web environment at $z\sim 1$}

\author[0009-0007-7887-783X]{Ivan Rapoport}
\email{ivanr@campus.technion.ac.il}
\affiliation{Physics Department, Technion -- Israel Institute of Technology, Haifa 3200003, Israel}

\author[0000-0003-2062-8172]{Vincent Desjacques}
\affiliation{Physics Department, Technion -- Israel Institute of Technology, Haifa 3200003, Israel}

\author[0000-0001-9735-4873]{Ehud Behar}
\affiliation{Physics Department, Technion -- Israel Institute of Technology, Haifa 3200003, Israel}

\author[0000-0002-2330-0917]{Ravi K. Sheth}
\affiliation{Center for Particle Cosmology, University of Pennsylvania, Philadelphia, PA 19104, USA}
\affiliation{The Abdus Salam International Center for Theoretical Physics, Strada Costiera 11, Trieste 34151, Italy}

\date{\today}

\begin{abstract}
Future near-infrared spectroscopic galaxy surveys will target high-redshift emission-line galaxies to test cosmological models. Deriving optimal constraints from emission-line galaxy clustering hinges on a robust understanding of their environmental dependence. Using the TNG300-1 simulation, we explore the correlation between properties of $\Halpha$ emitters and their environment anisotropy rather than traditional density-based measures. Our galactic $\Halpha$ emission model includes contributions from the warm interstellar medium. The environment anisotropy and type are assigned using a halo mass-dependent smoothing scale. We find that most luminous $\Halpha$ emitters ($L_{\Halpha}>10^{42} \ergss$) reside in filaments and knots. 
More generally, $\Halpha$ emitters are more biased in strongly anisotropic environments. While correlations with galactic properties are found to be weak, they are statistically significant for host halo masses $M\lesssim 10^{12}\ \hmsun$. Our analysis motivates further investigation into how environmental anisotropy influences galaxy evolution, and highlights the potential for leveraging these effects in the analyses of upcoming cosmological surveys. 
\end{abstract}

\section{Introduction}
\label{sec:intrp}

The next generation of galaxy surveys, including {\it Euclid} \citep{euclidcollaboration2024}, the Dark Energy Spectroscopic Instrument (DESI) \citep{desicollaboration2016}, the Spectro-Photometer for the History of the Universe, Epoch of Reionization and Ices Explorer (SPHEREx) \citep{spherexcollaboration2014} and the Nancy Grace Roman Space Telescope (RST) \citep{spergel2015} will provide detailed measurements of emission-line galaxy (ELG) clustering. These data will open new avenues for probing key cosmological scenarios, including the nature of dark matter and dark energy, the physics of the early Universe, and the mass of neutrinos. Interpreting the clustering and physical characteristics of ELGs requires a nuanced understanding of the environments in which they reside. Accurate modeling of ELG clustering further hinges on robust priors for galaxy bias parameters, as recent work has shown that the bias of ELGs is sensitive to both selection effects and the details of galaxy formation physics, motivating the use of simulation-calibrated priors and perturbative approaches \citep[e.g.,][for recent studies]{merson/etal:2019,barreira/etal:2020,zhai/etal:2021,barreira/etal:2021,yuan/etal:2022,marinucci/etal:2023,EuclidFlagship2,pei/etal:2024,madar/etal:2024,reyes/etal:2024,kazuyuki:2024,ivanov/etal:2024,DESI_ELG_1,DESI_ELG_2,martinez/etal:2025,ivanov:2025,Sullivan_2025}.

The cosmic environment plays a critical role in determining the properties and evolution of ELGs. The physical mechanisms responsible for their emission—such as intense star formation and active galactic nuclei (AGN) activity are intimately shaped by the exchange of gas between galaxies and their surroundings, linking ELG properties to the broader environment \citep[e.g.,][]{Kennicutt_1983,Keres2005,Hopkins_2012,Tumlinson_2017,nusser/etal:2020}. Consequently, ELGs are preferentially located in the dense structures of the cosmic web, especially filaments and knots \citep{Darvish_2014}. Recent work has shown that color-selected ELGs are more likely to reside in filamentary environments identified via smoothed density fields constructed from dark matter particle distributions \citep[][]{hadzhiyska/tachella/etal:2021}. 

Since there is no unique way to define the environment of galaxies, several methods have been developed to characterize it. Common approaches include local density estimators, such as counts-in-cylinders or adaptive smoothing techniques \citep[][]{Dressler1980,Cooper_2006} as well as classifications based on group membership or nearest-neighbor distances \citep[e.g., ][]{Baldry2006,Yang2009}. 
Beyond simple density-based measures \citep[e.g.][]{hoyle/etal:2005,park/etal:2007,moustakas/blanton:2009,tempel/saar/etal:2011}, the large-scale tidal field provides a comprehensive framework for characterizing the cosmic web  \citep[][]{Shen_2006,Hahn_2007,foreroromero/etal:2009,Libeskind2018,paranjape/etal:2018,Ramakrishnan_2019,alam/zu/etal:2019,favole/etal:2022,osato/okumura:2023}. In particular, tidal anisotropy, which quantifies the degree to which the local gravitational field is directionally dependent, encodes information about the geometry and dynamics of the matter distribution surrounding galaxies and halos. Tidal anisotropy has been shown to correlate strongly with halo assembly bias, concentration, shape, and large-scale bias \citep[][]{Ramakrishnan_2019}. 

The theoretical motivation for using the tidal field traces back to the nature of initial conditions. Perturbations in Gaussian random fields are inherently triaxial \citep{zeldovich:1970,doroshkevich:1970,bbks,heavens/peacock:1988,bond/myers:1996,sheth/mo/tormen:2001,jing/suto:2002}, and the large-scale shear field plays a central role in shaping the cosmic web from early-time Gaussian fluctuations to the highly nonlinear filamentary structure observed at late times \citep[e.g.][]{hoffman:1986,peebles:1990,dubinski:1992,bertschinger/jain:1994,bond/kofman/pogosyan:1996,sheth/mo/tormen:2001,desjacques:2008,pogosyan/etal:2009,hiding/etal:2014,aung/cohn:2016,musso/etal:2018,shim/etal:2021}.

Observationally, galaxy properties such as star formation rate and morphology are known to correlate with their cosmic web environment. For example, galaxies in filaments in the SDSS DR8 main galaxy sample tend to be less star-forming and exhibit earlier-type morphologies than those in more diffuse environments \citep{okane/etal:2024}. However, the correlation between large-scale environment and halo properties is subtle. While large-scale bias is known to correlate more strongly with local overdensity measures like the “halo-centric” $\delta_R$ than with halo mass \citep{abbas/sheth:2007,shi/sheth:2018,repp/szapudi:2022}, such density-based metrics are largely insensitive to the shape of the surrounding matter field—a limitation overcome by measures such as tidal anisotropy.

In this paper, we build upon these insights to explore further the interplay between the properties of high-redshift ($z\sim 1$) $\Halpha$ emitters (surveyed by Euclid, DESI, RST) and their environment. We use the $\Halpha$ emission line model presented in \cite{rapoport/etal:2025}, which includes not only the contribution from H II regions but also the $\Halpha$ emission arising from the warm diffuse interstellar and circumgalactic medium. On physical grounds, the fact that halos of mass $M\sim 10^{11}\ M_\odot/h$, which are the typical halos virializing at redshift $z\sim 1$, have a virial temperature $kT_\text{vir}\sim 12\ {\rm eV}$ of order the Ly$\beta$ transition energy strongly suggests that the warm phases contribute significantly to the $\Halpha$ flux of the $z\sim 1$ galaxies surveyed by Euclid and DESI. 

The paper is organized as follows: we briefly recap our model of galactic $\Halpha$ emission in section \S\ref{sec:Halpha}; describe ways of quantifying the environment in section \S\ref{sec:environment}; present our results for correlations of ELG properties and the environment in section \S\ref{sec:correlations}; and conclude in \S\ref{sec:conclusions}.

\section{Simulating $\Halpha$ emitters}
\label{sec:Halpha}

In \cite{rapoport/etal:2025}, we presented a physically-motivated model for galactic $\Halpha$ emission, which can be applied to hydrodynamical simulations of galaxy formation. Our model implements contributions to the $\Halpha$ luminosity induced by collisional excitation (CE), photo-excitation by the radiation flux of stars and active galactic nuclei (PE, AGN), radiative recombinations (RR), and H II regions (which are usually unresolved in hydrodynamical simulations) at the level of gas cells or particles.  This allowed us to produce spatially resolved $\Halpha$ emission maps.

\subsection{The $\Halpha$ emission line model}
\label{sec:sim}

The model assumes that the population of the atomic levels are in steady-state and can be computed within the coronal approximation where excitation occurs either from the HI ground state or by recombination of a bare (HII) proton. These approximations are justified by the relatively low densities and by the transparency of the interstellar medium (ISM). As a result, the total $\Halpha$ luminosity of each galaxy can be expressed as a sum over the independent contributions of each $\Halpha$ emission mechanism, i.e. 
\begin{equation}
    L_{\Halpha} =L_{\Halpha}^{\text{CE}}+L_{\Halpha}^{\text{PE}}+L_{\Halpha}^{\text{RR}}+L_{\Halpha}^{\text{AGN}}+L_{\Halpha}^{\text{HII}} \;.
\end{equation}
These contributions to the $\Halpha$ luminosity are computed from all the gas cells bound to the host subhalo. 

Beyond the local physical parameters obtained from the simulations, the key model parameters are:
\begin{itemize}
    \item The fraction $\lambda_h$ of gas in the hot phase for star-forming cells. 
    \item The characteristic distance $r_0$ for the absorption of Lyman photons in galaxies. 
    It is simultaneously used as the distance beyond which stars do not contribute to $L_{\Halpha}^{\text{PE}}$ 
    and the distance from the galactic center over which $L_{\Halpha}^{\text{AGN}}$ 
    is suppressed.
    \item The magnitude $C_{\text{HII}}$ of the contribution of H II regions to the $\Halpha$ luminosity.
\end{itemize}
While we set $\lambda_h=0.1$ for all star-forming cells as in \citet{rapoport/etal:2025}, we improve our original model in two ways.
First, we account for galaxy-to-galaxy variations in the photon mean free path according to the simple prescription
\begin{equation}
        \lambda_{\text{mfp}} = \min\left(r_0\left(\frac{R_{\text{gas}}}{10\ \text{kpc}}\right)^3\left(\frac{M_{\text{gas}}}{10^{10}M_\odot/h}\right)^{-1}Z^{-1},R_{\text{gas}}\right) \;,
    \end{equation}
where $R_{\text{gas}}$ is the gas half-mass radius, $M_{\text{gas}}$ is the total galactic gas mass, and $Z$ is the mass-weighted average gas metallicity, in solar units. 

Second, we explicitly model galactic $\Halpha$ dust extinction using a metallicity-dependent dust cross-section that links the gas content and metallicity of a galaxy to the dust attenuation of its $\Halpha$
emission. Further details will be presented in a forthcoming paper.
We fit $r_0$ and $C_{\text{HII}}$ to measurements of the $\Halpha$ luminosity functions at $z=1$, obtaining $r_0=0.72\ \rm{kpc}$ (physical) and $C_{\text{HII}}=10^{41.36}\ \left[\frac{\ergss}{M_\odot \text{yr}^{-1}}\right]$ . 
This calibration fully determines the model parameters. 
Note that the best-fit value of $r_0$ implies a cross section (per H-atom) $\sigma\sim 10^{-21}\ \rm{cm^2/H}$ consistent with typical dust-absorption cross sections for UV/Lyman-$\alpha$ 
radiation \citep{Draine2011}.

\subsection{Numerical simulations}
\label{sec:sim}

We focus on the $z=1$ snapshot extracted from the TNG300-1 simulation \citep{TNG_DR,TNG_1,TNG_2,TNG_3,TNG_4,TNG_5}, for which the adopted cosmology is $\Omega_m=0.3089,\ \Omega_\Lambda =0.6911,\ \Omega_b=0.0486,\ h=0.6774 $ for the matter, dark energy and baryonic energy densities and the Hubble expansion rate (in units of $100\hkmsmpc$), respectively. Dark matter parent halos are identified by a standard friends-of-friends (FoF) algorithm with a linking length of 0.2 in units of the mean interparticle spacing. Our fiducial simulation is TNG300-1, which has a box size $L=205 \hmpc$ and contains $2500^3$ dark matter particles and an equal number of gas particles. We use the higher resolution TNG50-1 \citep{TNG50a,TNG50b}, which evolves $2\times 2160^3$ dark matter and gas particles in a box of size $L=35 \hmpc$, to assess the impact of numerical resolution on the correlations reported in this work. We incorporate our $\Halpha$ emission line model into all the subhalos resolved in the simulations.
Throughout the paper, we identify $\Halpha$ emitters as suhalos with resulting luminosity $L_{\Halpha}>10^{39}\ergss$. 
For brevity, we will also henceforth refer to them as ELGs. 

\subsection{Halo Occupation Distribution}
\label{sec:hod}

\begin{figure}
    \centering
    \includegraphics[width=0.4\textwidth]{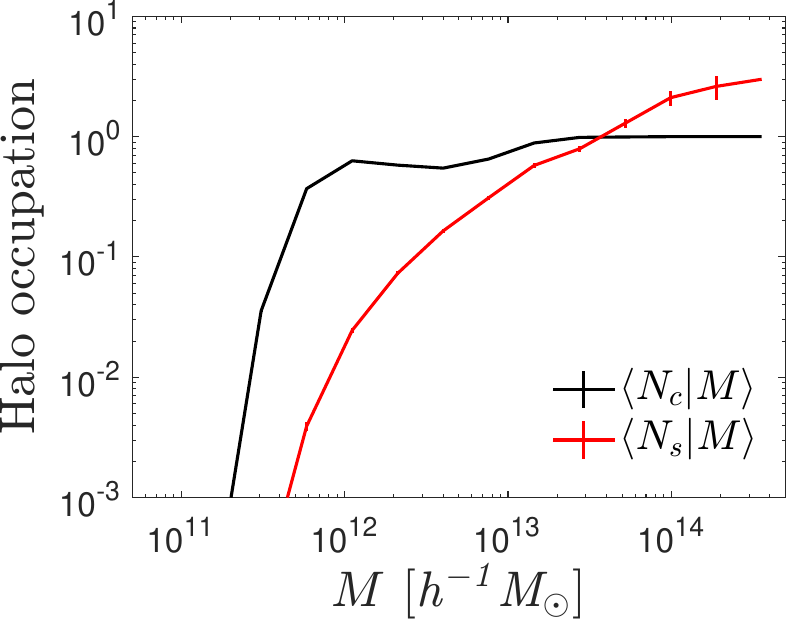}
    \caption{The average central and satellite occupation numbers $\langle N_c|M\rangle$, $\langle N_s|M\rangle$ computed for our $\Halpha$ emission model for TNG300-1 at $z=1$. }
    \label{fig:HOD}
\end{figure}

The TNG suite of simulations includes information on whether an ELG is a central or a satellite galaxy. 
This allows us to compute the halo occupation distribution (HOD) of central and satellite ELGs, $N_c$ and $N_s$, as a function of the host halo mass $M$.
The HOD framework provides a fast and flexible approach to modelling galaxy clustering \cite[see][for a review]{cooray/sheth:2002}. 
For illustration, we show in Fig.~\ref{fig:HOD} the HOD of ELGs identified in the $z=1$ snapshot of TNG300-1. 
The average central occupation number $\langle N_c|M\rangle$ approaches unity at high halo mass and exhibits a local peak at $\sim 10^{12}M_\odot/h$. 
The average satellite occupation number $\langle N_s|M\rangle$ increases with $M$, with a shallower rise for rich groups and clusters ($M\gtrsim 10^{14}\hmsun$). The fraction of halos which host a satellite ELG when the central galaxy is not an ELG is small ($<0.01\%$).

\section{Quantifying the galactic environment}
\label{sec:environment}


\subsection{Tidal and strain rate tensor}
\label{sec:T_V_anisotropy}

We use the gravitational tidal and strain rate (velocity shear) tensors, $\Psi_{ij}$ and $\Sigma_{ij}$, to quantify the anisotropy of the environment. The gravitational tidal tensor is defined as
\begin{equation}
    \Psi_{ij}=\partial_i\partial_j \phi \;,
\end{equation}
where $\phi(\Vec{x})$ is a potential obtained by solving the normalized Poisson equation
\begin{equation}
    \nabla^2 \phi = \delta \;,
\end{equation}
where $\delta(\Vec{x})$ is the dark matter overdensity field. We note that trivially, $\text{Tr}(\Psi_{ij})\equiv\delta$. The strain rate tensor is defined as
\begin{equation}
    \Sigma_{ij}=-\frac{1}{2aHf}\left(\partial_iv_j + \partial_jv_i\right) \;,
\end{equation}
where $\Vec{v}(\Vec{x})$ is the dark matter velocity field, $H(a)$ is the Hubble parameter at scale factor $a$ and $f\approx (\Omega_m(a))^{0.55}$ is the logarithmic growth rate. $\Sigma_{ij}$ is dimensionless, and this choice of scaling ensures $\Psi_{ij}$ and $\Sigma_{ij}$ are equal in linear theory. The minus sign in the definition of $\Sigma_{ij}$ makes positive eigenvalues of the tensor denote contraction/compression of matter along corresponding eigenvectors. Note that $\Psi_{ij}$ and $\Sigma_{ij}$ generally differ beyond linear theory. In what follows, the subscript T and V will denote quantities computed via the tidal tensor or the strain rate tensor, respectively.

\subsection{Anisotropy parameter $\alpha$}
\label{sec:alpha}

To quantify the anisotropy of the environment around halos and galaxies, we first compute $\Psi_{ij}$ and $\Sigma_{ij}$ in the simulation volume. In practice, we interpolate the dark matter particles on a regular cubical grid of size $N^3$ using a cloud-in-cell (CIC) assignment to construct the density and velocity fields $\rho$ and $\Vec{v}$. We have checked that accounting for baryonic matter does not lead to notable changes in our results. Next, we Fourier transform $\rho$ and $\Vec{v}$ and compute the Fourier modes of $\Psi_{ij}$ and $\Sigma_{ij}$, which we correct for the CIC assignment and smooth further with a Gaussian low-pass filter $\exp (-|\Vec{k}|^2\rm{R_G^2}/2)$ on scale $\rm{R_G}$. We vary ${\rm R_G}$ in the range $0.25\lesssim \rm{R_G}\lesssim 2 \hmpc$ to obtain, after a backward Fourier transform, measurements of the smoothed $\Psi_{ij}(\Vec{x};{\rm R_G})$ and $\Sigma_{ij}(\Vec{x},{\rm R_G})$ on the cubical grid for a range of ${\rm R_G}$, where
\begin{equation}
\label{eq:RGvsM}
\text{R}_{\text{G}}(M)=4\text{R}_{\text{vir}}(M)/\sqrt{5}
\end{equation}
as advocated in \cite{Ramakrishnan_2019}. Here, $\text{R}_{\text{vir}}(M)$ is the virial radius of a halo of mass $M\equiv M_{\text{200b}}$. 
This procedure should be accurate so long as $\text{R}_{\text{G}} \gtrsim L/N$, where $L$ is the box side length. Hence, for TNG300-1 a grid of size $N^3=800^3$ is sufficient to resolve halos down to a halo mass $M\approx 2\times 10^{11}M_\odot/h$.

\begin{figure}
    \centering
    \includegraphics[width=0.45\textwidth]{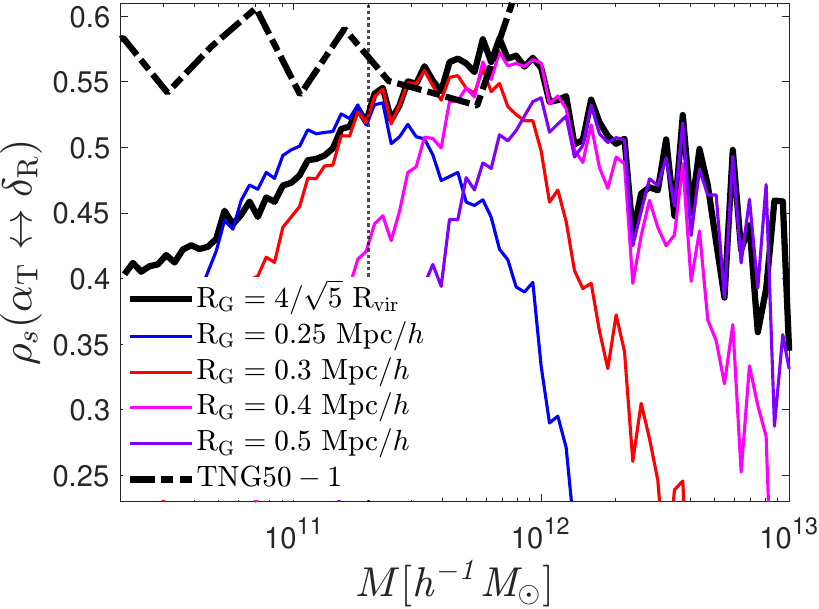}
    \caption{Spearman rank correlation coefficient $\rho_s$ between the anisotropy parameter $\alpha_{\rm T}$ and the smoothed overdensity $\delta_{\rm R}$ at the location $\Vec{x}_h$ of dark matter halos for Gaussian filters with different prescriptions of the smoothing radius $\text{R}_\text{G}$. Results are shown as a function of the halo mass $M$. The dot-dashed line shows the correlation extracted from TNG50-1. The thin vertical line marks the minimum mass $M= 2\times 10^{11}\hmsun$ below which $\rm{R_G}$ is poorly resolved with TNG300-1 (see text).}
    \label{fig:rho_S_alpha_delta}
\end{figure}

The dimensionless anisotropy parameters $\alpha_{\rm{T}}(\Vec{x};\rm{R_G})$ and $\alpha_{\rm{V}}(\Vec{x};\rm{R_G})$ are computed on the cubical grid from the eigenvalues $\lambda_1, \lambda_2, \lambda_3$ of the tensors $\Psi_{ij}(\Vec{x};{\rm R_G}(M))$ and $\Sigma_{ij}(\Vec{x};{\rm R_G}(M))$ following \cite{paranjape/etal:2018} and \cite{Ramakrishnan_2019},
\begin{equation}
    \alpha(\Vec{x};\text{R}_{\text{G}}) = \frac{\sqrt{\frac{1}{2}\left[(\lambda_2-\lambda_1)^2+(\lambda_3-\lambda_1)^2+(\lambda_3-\lambda_2)^2\right]}}{1+\lambda_1+\lambda_2+\lambda_3} \;.
\end{equation}
Finally, the anisotropy parameters defined on a grid are interpolated in space to the center-of-mass (CoM) positions $\Vec{x}_h$ of halos, and in $\rm{R_G}$ to the halo masses $M$, to produce halo-centric anisotropies $\alpha_{\rm{V}}$, $\alpha_{\rm{T}}$. Halos with a low $\alpha$ ($\alpha\lesssim 0.2$) are typically surrounded by a fairly isotropic environment, while those with a large $\alpha$ ($\alpha\gtrsim 0.5$) usually reside in anisotropic filamentary environments. \cite{paranjape:2021} provides theoretical insights into the behaviour of $\alpha$.
Note that, by the definition of $\Psi_{ij}$ and $\Sigma_{ij}$, $\rm{\alpha_T}$ and $\rm{\alpha_V}$ are equal in linear theory.
Finally, we assign to each galaxy --- whether central or satellite --- the anisotropy parameters of the host parent halo.

As a sanity check, we have measured the strength of the correlation $\alpha_{\rm{T}} \leftrightarrow \delta_{\rm{R}}$ as a function of halo mass. Here, $\delta_{\rm R}$ is the overdensity smoothed on scale $\text{R}_\text{G}$, which is varied as in Eq.~\eqref{eq:RGvsM}. The Spearman rank order correlation coefficient $\rho_s$, which quantifies monotonic relationships between two variables regardless of their specific functional form, is shown in Fig.~\ref{fig:rho_S_alpha_delta} and reproduces the findings of \cite{Ramakrishnan_2019}. Namely, the correlation is always positive and, at fixed halo mass, the smoothing prescription of Eq.~\eqref{eq:RGvsM} maximizes the correlation strength, which reaches a maximum at $M\sim 10^{12} \hmsun$. The dotted vertical line indicates the minimum mass $M\sim 2\times 10^{11}\hmsun$ below which $\rm{R_G}$ is smaller than the grid spacing $L/N$ in TNG300-1. To determine whether the correlation persists below this mass scale, we have extracted $\rho_s$ from TNG50-1 and found that its magnitude is comparable to the maximum value of $\rho_s$ measured ins TNG300-1. This suggests that the decline of $\rho_s$ below $M\sim 10^{12}\hmsun$ in TNG300-1 is likely a consequence of the finite grid resolution.

\begin{figure}
    \centering

\includegraphics[width=0.44\textwidth]{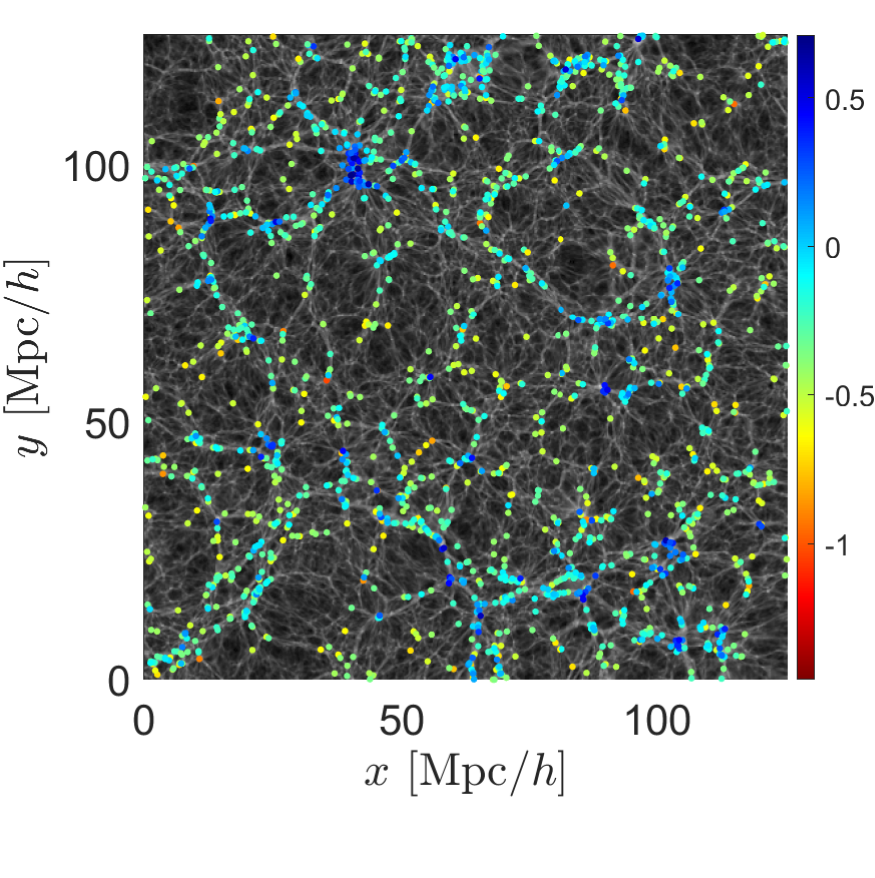}
    \caption{A two-dimensional slice extracted from the $z=1$ snapshot of TNG300-1. The points indicate the positions of halos with $M>2\times 10^{11}\hmsun$, and are colored progressively according to their $\log_{10} \alpha_{\text{T}}$, as indicated by the color bar. The background gray scale shows the dark matter overdensity $\log \ (\rho_{\text{DM}}/\bar{\rho}_{\text{DM}})$.}
    \label{fig:ELG_cosmic_web}
\end{figure}

Now that we are confident that our estimates of $\alpha$ are reasonable, Fig.~\ref{fig:ELG_cosmic_web} shows the distribution of halos having masses in excess of $2\times 10^{10}\hmsun$, each colored by its value of $\alpha_{\text{T}}$, measured in a two-dimensional slice extracted from the $z=1$ snapshot of TNG300-1.  The underlying black-and-white cosmic web indicates the dark matter distribution in the snapshot. This provides a visual impression of the sensitivity of $\alpha$ to the large scale clustering of mock galaxies.  

\begin{figure}
    \centering
    \includegraphics[width=0.32\textwidth]{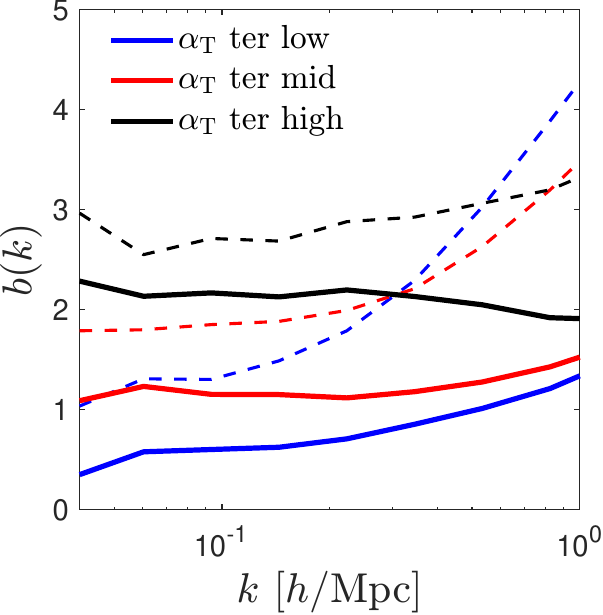}
    \includegraphics[width=0.32\textwidth]{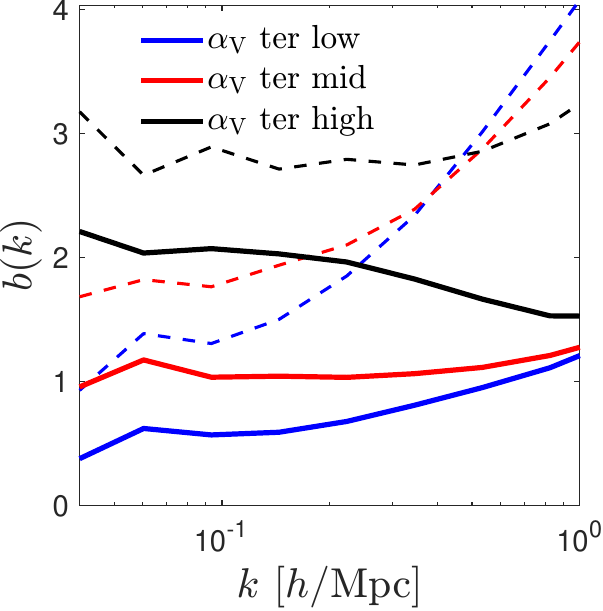}
    \caption{{\bf Top}: Measurements of the scale-dependent bias $b(k)$ of mock ELGs with $\Halpha$ luminosity $L_{\Halpha}>10^{41}\ergss$ when split into terciles of $\rm{\alpha_T}$. Results are shown separately for central (solid curves) and satellite galaxies (dashed curves). {\bf Bottom}: Same as the top panel but for $\rm{\alpha_V}$.}
    \label{fig:ELG_bias_terciles}
\end{figure}

To quantify this, 
we have measured the scale-dependent bias "transfer function" 
\begin{equation}
    b(k) = \frac{P_{\text{gm}}(k)}{P_{\text{mm}}(k)}
\end{equation}
for mock ELGs with luminosity $L_{\Halpha}>10^{41}$ and $L_{\Halpha}>10^{42}\ergss$. 
$b(k)$ is a measure of the clustering of galaxies relative to the matter distribution.
The galaxy-matter cross-spectrum $P_{\rm gm}(k)$ and the matter power spectrum $P_{\rm mm}(k)$ are computed on the $800^3$ cubical grid.
Results are shown in Fig.~\ref{fig:ELG_bias_terciles} for central and satellite galaxies as the solid and dashed curves, respectively. They have been split into terciles of $\alpha_{\rm T}$ and $\alpha_{\rm V}$ of their parent halos as indicated in the figure.

Although the simulation is not large enough to capture the convergence of $b(k)$ to a constant $b_1$ on linear scales, ELGs residing in tidally sheared (high $\alpha$) parent halos are clearly more biased. In general, $b(k)$ of the satellite galaxies is larger than that of central galaxies because the former preferentially reside in massive dark matter halos (c.f. Fig.~\ref{fig:HOD}). The scale dependence of $b(k)$ becomes apparent for $k\gtrsim 0.1\hmmpc$.  While it is different for central than for satellite galaxies, our results consistently show that samples with the lowest values of $\alpha_{\rm T}$ or $\alpha_{\rm V}$ exhibit the strongest positive upturn in $b(k)$ -- i.e. most scale dependent bias. This is somewhat at odds with the findings of \cite{Ramakrishnan_2020} (i.e., halos with larger values of $\alpha$ have larger values of $b_2$) unless bias parameters other than $b_2$ contribute significantly to the scale-dependence of $b(k)$ for $k\gtrsim 0.1\hmmpc$.

\begin{figure*}
    \centering
    \includegraphics[width=0.42\textwidth]{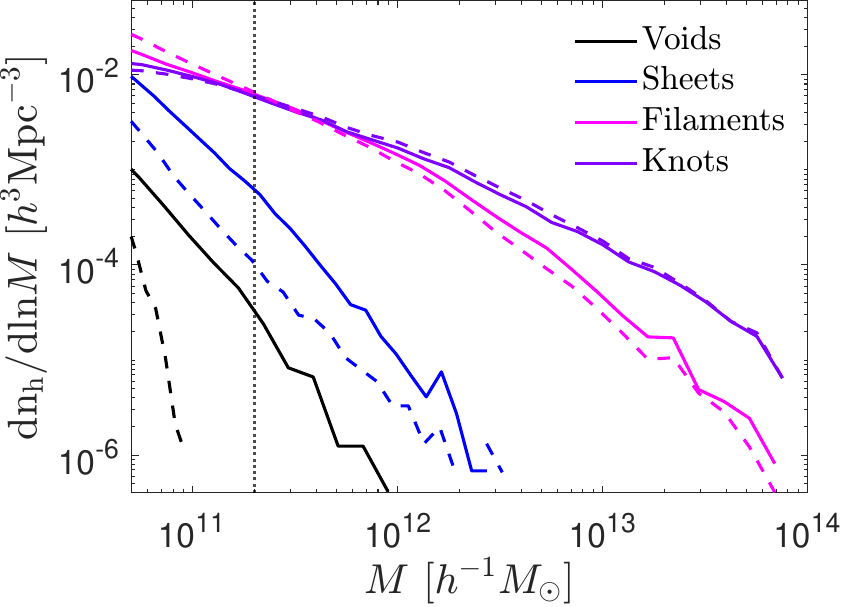}
    \includegraphics[width=0.4\textwidth]{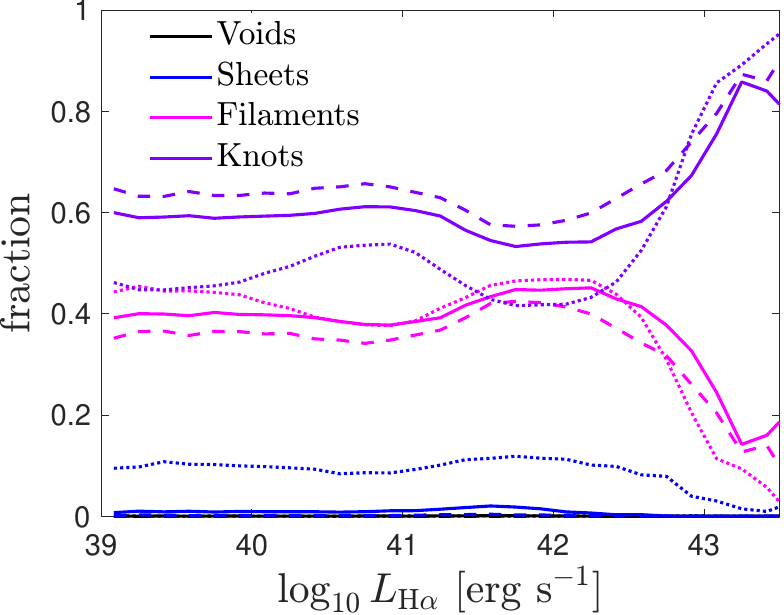}
    \caption{\textbf{Left panel:} Comoving differential halo mass function, conditioned on environment. Solid and dashed curves indicate the classification obtained from the V- and T-tensors, respectively. Vertical line indicates, as in Fig.~\ref{fig:rho_S_alpha_delta}, the mass $M\sim 2\times 10^{11}\hmsun$ below which the smoothing scale $\rm{R_G}$ is poorly resolved. \textbf{Right panel:} 
    Fraction of ELGs in voids, sheets, filaments and knots as a function of $\Halpha$ luminosity (color and linestyle conventions same as left panel). 
    The classification obtained from DisPerSE (dotted) predicts fewer ELGs in knots and more in sheets. Due to the limited resolution of the grid, ELGs with parent halo mass smaller than $2\times 10^{11} \hmsun$ are not included here.}
    \label{fig:classification_halo}
\end{figure*}

\subsection{Cosmic web classification}
\label{sec:cosmicweb}

Halo-centric measurements of $\Psi_{ij}$ and $\Sigma_{ij}$ can be used to classify the environment of dark matter halos and the ELGs they host. Classifications based on $\Psi_{ij}$ and $\Sigma_{ij}$ are known as T-web and V-web respectively \citep[see][for detailed discussions]{Hahn_2007,Hoffman_2012,Libeskind_2012,Pfeifer_2022}.
We define $\lambda_{\text{th}}$ to be the eigenvalue threshold above which the local matter distribution is considered as collapsing along the corresponding principal axis. A value of $\lambda_\text{th}\sim \mathcal{O}(1)$ usually returns a
visual impression of the cosmic web close to that observed in actual data \citep{Hoffman_2012}.
Since $\Psi_{ij}$ and $\Sigma_{ij}$ differ in the nonlinear regime, we set $\lambda_{\text{th,V}}=0.4 H_0/(aHf)\approx 0.515$ and $\lambda_{\text{th,T}}=1.5 H_0/(aHf)\approx 1.932$ for the strain rate and tidal tensor at $z=1$, respectively. The choice of $\lambda_{\text{th,V}}$ is consistent with
\cite{foreroromero/etal:2009,martizzi/etal:2019,Pfeifer_2022}, who advocated $\lambda_\text{th}=0.2 - 0.4$. Our values of $\lambda_\text{th,T}$, $\lambda_\text{th,V}$ are chosen such that the V- and T-web yield a similar cosmic web classification. 

The eigenvalues of each tensor are computed on a grid following the procedure described in \S\ref{sec:T_V_anisotropy}. They are interpolated in configuration space at the halo CoM position, and in $\rm{R_G}$-space according to the halo mass $M$. Finally, we assign an environment mark to each halo depending on the number $N_\lambda$ of eigenvalues above the threshold $\lambda_{\text{th}}$, with $N_\lambda=0$, 1, 2 and 3 corresponding to voids, sheets, filaments and knots respectively.

We noted above that halos with large $\alpha_{\rm T}$ or $\alpha_{\rm V}$ are more strongly clustered (c.f. Fig.~\ref{fig:ELG_bias_terciles}).  
Clustering strength is usually thought to be strongly correlated with halo mass, so it is important to separate the effects of environment from those of halo mass.  To set the stage, 
the left panel of Fig.~\ref{fig:classification_halo} shows that the most massive halos tend to occur in knots or filaments.  The abundance of halos with $M\ge 10^{12}\hmsun$ is lower by more than an order of magnitude in sheets and voids.  In addition, notice that, as mass increases, these abundances decrease more steeply for sheets than for knots or filaments.  This finding is the same for both the V- and T-web classifications and will be relevant in the next section, when we discuss spatial clustering.  

One can use the topological representation of the cosmic web implemented in DisPerSE \citep{DisPerSE} to assign an additional environmental mark. DisPerSE estimates the density field using the discrete halo or galaxy distribution and identifies critical points. Local maxima and minima of the density field define the knots and voids of the cosmic web, respectively. Furthermore, filaments correspond to the unique integral lines connecting knots, while sheets are associated with minima along filaments. 
DisPerSE classifications are publicly available for galaxies in the TNG300-1 simulation, so we use them to assign an environment to each galaxy.

The right panel of Fig.~\ref{fig:classification_halo} displays the environmental mark fractions of ELG parent halos as a function of the ELG total $\Halpha$ luminosity. In each luminosity bin, we computed the fraction of galaxies in every classification, such that the fractions sum to 1 by definition. The threshold values $\lambda_\text{th,V}$ and $\lambda_\text{th,T}$ adopted here ensure that the V- and T-web classifications are in reasonably good agreement with those predicted by DisPerSE for $\Halpha$ luminosities greater than $10^{41}\ergss$. Observe also that the fraction of ELGs in filaments is nearly constant and equal to $\approx 40\%$ across four orders of magnitude in luminosity, unlike the fraction of galaxies in sheets. At $L_{\Halpha}>10^{43} \ergss$, nearly all galaxies are in knots.   
Finally, note that ELGs with parent halo mass smaller than $2\times 10^{11} \hmsun$ are not included in this classification due to the limited resolution of the grid. 
This excludes only $\approx 18\%$ ($2\%$) of ELGs with $L_{\Halpha}>10^{39}\ergss$ ($L_{\Halpha}>10^{42}\ergss$). 

\subsection{Environmental bias}
\label{sec:envbias}

The peak-background split argument relates the magnitude of the large scale bias $b_1$ to (minus) the slope of the halo mass function \citep{kaiser:1984,bbks,cole/kaiser:1989,mo/white:1996,sheth/tormen:1999}.
In light of this, the left panel of Fig.~\ref{fig:classification_halo} suggests that, at fixed halo mass, halos in sheets and filaments should be significantly more clustered than those residing in knots. 

\begin{figure}
    \centering
    \includegraphics[width=0.39\textwidth]{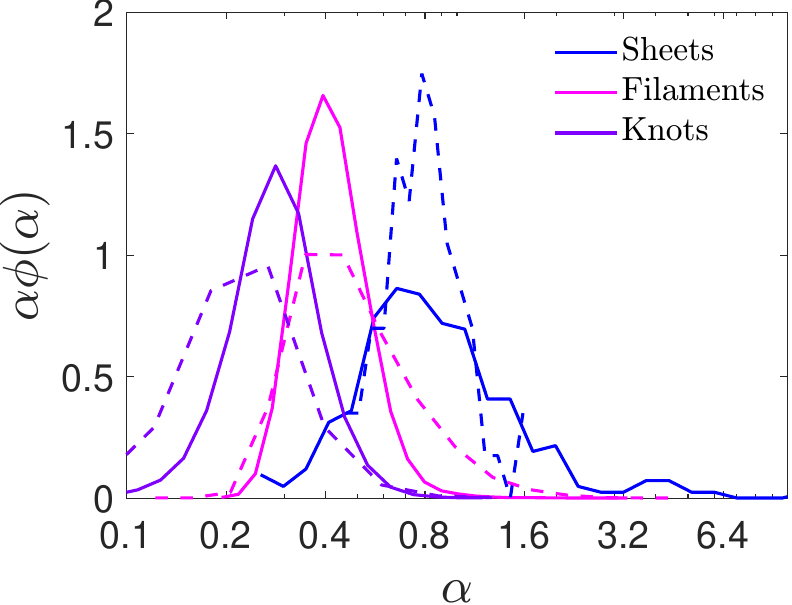}
    \includegraphics[width=0.4\textwidth]{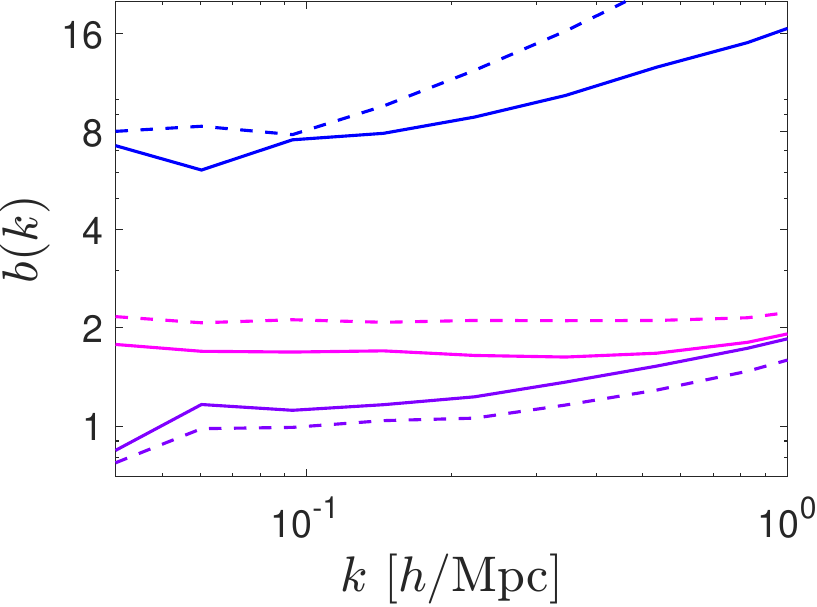}
    \caption{\textbf{Top:} The probability distribution function (PDF) of $\rm{\alpha_V}$ (solid) and $\rm{\alpha_T}$ (dashed) for halos with $M>5\times 10^{11} M_\odot/h$ when conditioned on environment (as determined by the V- and T- tensors, respectively), with the same color and linestyle convention of Fig.~\ref{fig:classification_halo}.  \textbf{Bottom:} The corresponding scale-dependent bias transfer functions $b(k)$.}
    \label{fig:env_halo_alpha_bias}
\end{figure}

To illustrate this point, the top panel of Fig.~\ref{fig:env_halo_alpha_bias} shows the distribution of the anisotropy parameters $\alpha_{\rm V}$ and $\alpha_{\rm T}$ for halos with mass $M\geq 5\times 10^{11}\ M_\odot/h$. The anisotropy parameter thus is a good discriminator of the environment. Furthermore, the largest $\alpha_{\rm V}$ and $\alpha_{\rm T}$ values are exclusively associated with the rare halos residing in sheets. 
The bottom panel of Fig.~\ref{fig:env_halo_alpha_bias} displays the corresponding bias transfer functions $b(k)$ conditioned on halo environment.  It demonstrates that at fixed halo mass, halos in sheets cluster most strongly, even though halos in knots and filaments tend to be more massive on average. In the T-web classification, the mean halo masses are $\approx 10^{10.4}$, $10^{10.5}$, $10^{11.1}$ and $10^{11.6} \hmsun$ for halos in voids, sheets, filaments and knots, respectively. The difference in the clustering of halos residing in filaments and in knots is also significant. These findings are consistent with previous work showing the strong dependence of halo abundances and clustering on $\alpha$ \citep[see][]{paranjape/etal:2018}.

\begin{figure*}
    \centering
    \includegraphics[width=0.43\textwidth]{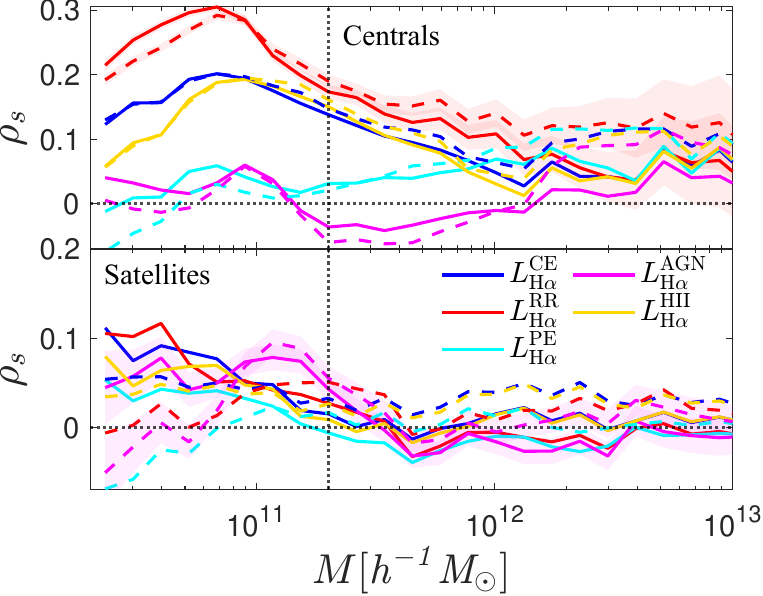}
    \includegraphics[width=0.45\textwidth]{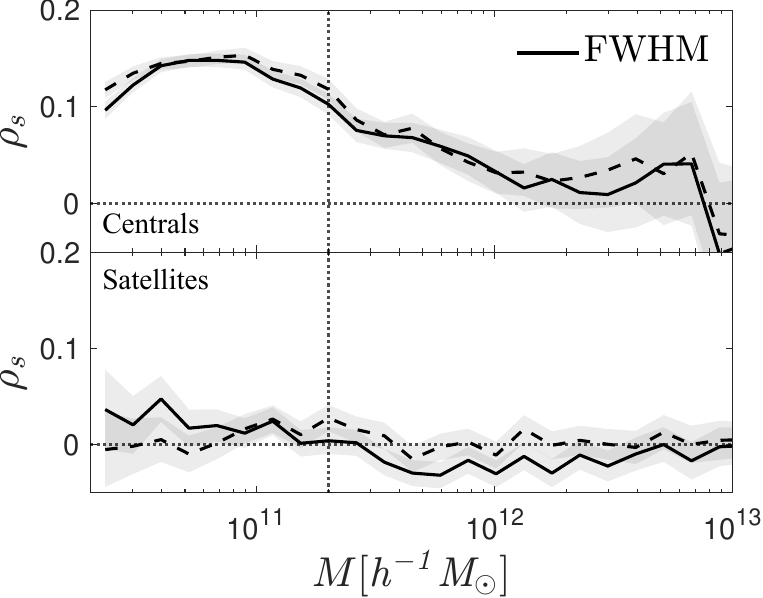}
    \caption{\textbf{Left:} Spearman rank order coefficient $\rho_s$ between the anisotropy parameter and sources of $\Halpha$ emission as a function of the mass $M$ of the parent host halo. Solid (dashed) curves show the correlations with $\alpha_{\rm{V}}$ ($\alpha_{\rm{T}}$). \textbf{Right:} Correlation coefficient between the anisotropy parameter and the line width of the $\Halpha$ emission line. The upper panels display the results for central galaxies, and the lower panels for satellites. The dotted horizontal lines indicate $\rho_s=0$ while the vertical line indicates, as in Fig.~\ref{fig:rho_S_alpha_delta}, the mass $M\sim 2\times 10^{11}\hmsun$ below which the smoothing scale $\rm{R_G}$ is poorly resolved.}
    \label{fig:rho_S_alpha_LHa}
\end{figure*}

\section{Correlation between ELG properties and environment}
\label{sec:results}

Having established that the large-scale bias of ELGs strongly correlates with their anisotropy parameter, we now quantify in this section how strongly anisotropy correlates with galaxy properties.

\subsection{ELG properties}
\label{sec:galproperties}

We will focus on a few galaxy properties (loosely labelled by $I_g$) and examine their correlation with the anisotropy parameters $\alpha_{\rm{T}}$ and $\alpha_{\rm{V}}$ evaluated on the host halo scale. In addition to the $\Halpha$ luminosity, we will focus on 5 properties: the instantaneous SFR (star formation rate), the total galactic gas mass $m_{\rm{gas}}$, the 1-dimensional velocity dispersion $\sigma_V$ of all particles/cells (i.e. gas, stars, dark matter) bound to the galaxy, the total acccretion rate $\dot{M}_{\rm{BH}}$ of all black holes in the galaxy, and the total galactic gas inflow rate $\dot{m}_{\rm{gas}}$. The first four quantities can be extracted directly from the public galaxy catalogs\footnote{\texttt{SubhaloMassType} for the gas mass, \texttt{SubhaloVelDisp} for the velocity dispersion, \texttt{SubhaloBHMdot} for the black hole accretion rate and \texttt{SubhaloSFR} for the total SFR.} whereas, for the instantaneous gas inflow rate, we use the expression adopted by~\cite{Nelson_2019}. Namely, we determine the rate of change in the total galactic gas mass from the Lagrangian properties of gas cells at a given time according to
\begin{equation}
    \label{eq:m_dot_Nelson}
    \dot{m}_{\text{gas}}=-\frac{1}{\Delta r}\sum_{\substack{\text{gas cells} \\ |r-r_0|<\Delta r/2}} \frac{\Vec{v}\cdot \Vec{r}}{r}m_{\text{gas,cell}} \;,
\end{equation}
which, up to sign, is a discretized form of the expression $\dot{m}=\int_S \rho(\Vec{v}\cdot \hat{n}) dS$ that encodes conservation of mass, where $S$ is some 2-dimensional surface with perpendicular unit vector $\hat{n}$; $\rho,\Vec{v}$ are the gas density and velocity respectively.
Here, the sum runs over all the gas cells bound to the galaxy, which lie in a spherical shell of radius $r_0$ and width $\Delta r$ centered on the galaxy position (defined by the position of the most strongly bounded particle). Furthermore, $\Vec{r}$ and $\Vec{v}$ are the physical position and velocity of the particles (relative to the galaxy frame), $r=|\Vec{r}|$, and $m_{\text{gas,cell}}$ is the gas cell mass. Therefore, a positive value of $\dot{m}_{\text{gas}}$ indicates a net inflow of gas inside the shell of radius $r_0$. Note that we scale $r_0$ with the galaxy (total) half-mass radius $r_{1/2}$ to account for variations in the size of galaxies. For each galaxy, we compute three gas inflow rates $\dot{m}_{\text{gas}}^{(1)}$, $\dot{m}_{\text{gas}}^{(2)}$ and $\dot{m}_{\text{gas}}^{(3)}$ on assuming $r_0=2r_{1/2}$, $4r_{1/2}$ and $7r_{1/2}$, respectively. The width is scaled according to $\Delta r=r_0/4$. 

\subsection{Correlation with environment}
\label{sec:correlations}

\begin{figure}
    \centering
    \includegraphics[width=0.43\textwidth]{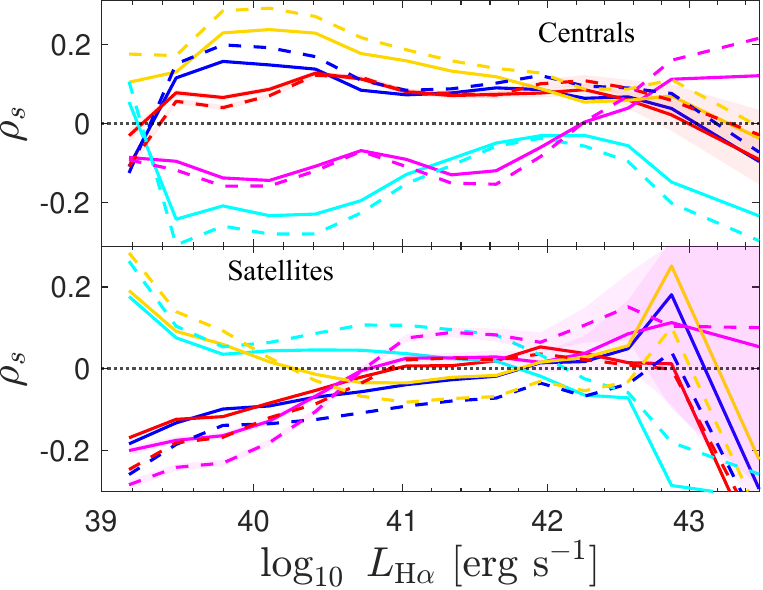}
    \caption{Same as the left panel of Fig.~\ref{fig:rho_S_alpha_LHa}, but shown as a function of the total $\Halpha$ luminosity of the mock ELGs.}
    \label{fig:rho_S_alpha_LHa_bins_lumi}
\end{figure}

We carry out a correlation analysis to identify the galaxy properties that correlate most with the anisotropy parameter. Instead of performing a Pearson test (which assumes a linear correlation), we use again the Spearman rank order correlation coefficient $\rho_s $ to test for any monotonic relation. 
Concretely, we compute $\rho_s(\alpha\leftrightarrow I_g)$, where $I_g$ is one of the galaxy properties outlined above. A positive $\rho_s(\alpha\leftrightarrow I_g)$ implies that $I_g$ tends to increase with $\alpha$ or, according to the top panel of Fig.~\ref{fig:env_halo_alpha_bias}, as one moves from knots to filaments and sheets.
Notice that the null probability decreases with increasing value of $|\rho_s|$ and sample size $N_s$.

The left panel of Fig.~\ref{fig:rho_S_alpha_LHa} shows the correlation coefficient $\rho_s$ between the anisotropy parameter and the different sources of $L_{\Halpha}$ luminosity in our emission line model. The solid and dashed curves were obtained with $\alpha_{\rm V}$ and $\alpha_{\rm T}$, respectively. 
They are shown as functions of halo mass for central (top panel) and satellite galaxies (bottom panel). 
The shaded areas show the uncertainty $\delta\rho_s$ on the measurement of $\rho_s$. It is computed according to the formula $\delta\rho_s = \sqrt{\frac{1-\rho_s^2}{N_s-2}}$. To avoid clutter, error bars are shown for only one source of $\Halpha$ emission.

The correlations weaken as $M$ increases significantly above $M_*$. For $M\sim M_\star$, $\alpha$ correlates most strongly with $L_{\Halpha}^{\rm RR}$ for central galaxies, whereas for satellite galaxies, the strongest correlation is with $L_{\Halpha}^{\rm AGN}$. The correlation of $\alpha$ and $L_{\Halpha}^{\rm AGN}$ indicates that satellite galaxies are influenced by AGN activity in nearby centrals
(AGN are mostly found in central galaxies).
This influence appears to suppress star formation in small satellite ELGs more efficiently in large $\alpha$ environments.

The correlation between $\alpha$ and $L_{\mathrm{H}\alpha}^{\mathrm{RR}}$ is consistent with a physical picture in which galaxies embedded in sheets and filaments experience large-scale accretion shocks. Such shocks are theoretically expected \citep{Mo2010Galaxy} and are observed in cosmological simulations \citep{mandelker/etal:2019,ramsoy/etal:2021,pasha/etal:2023}. They can influence galaxy evolution by heating and ionizing the circumgalactic gas and thus quenching star formation \citep{hasan/etal:2023,hasan/etal:2024}. In the TNG galaxy formation model, ELGs residing in low-mass halos within these environments accrete gas from a hot, ionized phase that produces $\mathrm{H}\alpha$ emission through radiative recombinations as it cools and condenses into the ISM. 
Therefore, we interpret the $\alpha \leftrightarrow L_{\mathrm{H}\alpha}^{\mathrm{RR}}$ trend as a manifestation of this environment-driven accretion regime, with the caveat that this is a simulation-informed scenario rather than a model-independent prediction.

The right panel of Fig.~\ref{fig:rho_S_alpha_LHa} shows the correlation between $\alpha$ and the $\Halpha$ line full width at half maximum (FWHM; see \S\ref{appendix:FWHM} for details of its computation) for central (top panel) and satellite galaxies (bottom panel). For centrals, a weak but non-vanishing correlation is detected up to $M\sim 10^{12}\ M_\odot/h$. 
In Appendix \S\ref{appendix:TNG50}, we have checked the numerical convergence of several of these correlations and found them to be robust down to a mass $M\sim {\rm a ~few}\ 10^{10}\hmsun$. 

Fig.~\ref{fig:rho_S_alpha_LHa_bins_lumi} is identical to Fig.~\ref{fig:rho_S_alpha_LHa} except that $\rho_s$ is displayed as a function of the total $\Halpha$ luminosity of the mock ELGs. It is interesting that the trends, when shown as a function of $L_{\Halpha}$, can be very different from those shown as a function of $M$. Consider $L_{\Halpha}^\text{RR}$ for instance. In Fig.~\ref{fig:rho_S_alpha_LHa}, the corresponding $\rho_s$ can be as large as 0.3 for $M\sim 10^{11}\ M_\odot/h$ whereas, in Fig.~\ref{fig:rho_S_alpha_LHa_bins_lumi}, $|\rho_s|$ does not exceed 0.1, regardless of the value of $L_{\Halpha}$. 
By contrast, plotting $\rho_s(\alpha\leftrightarrow L_{\Halpha}^\text{PE})$ as a function of $L_{\Halpha}$ reveals a negative correlation at low luminosities where PE dominates \citep{rapoport/etal:2025}. This trend is consistent with the physical picture outlined above: star formation is quenched in filamentary environments, which suppresses the PE rate.

\begin{figure}
    \centering
    \includegraphics[width=0.4\textwidth]{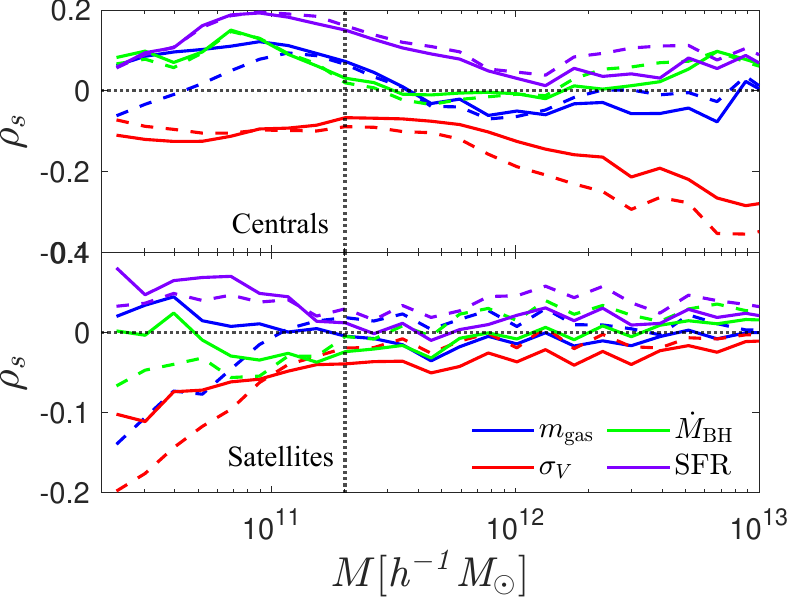}
    \caption{Correlations between $\alpha$ and a range of galaxy properties defined in \ref{sec:galproperties}: the total gas mass, 1-D velocity dispersion (all galaxy member particles), the BH accretion rate, the total stellar mass and the total star formation rate. The solid lines show correlations with $\alpha_{\rm{V}}$ while the dashed show $\alpha_{\rm{T}}$. The top panel shows correlations for central galaxies, while the bottom panel shows for satellite galaxies}
    \label{fig:rho_S_alpha_gal_params}
\end{figure}

Likewise, Fig.~\ref{fig:rho_S_alpha_gal_params} displays the Spearman rank order correlation coefficient $\rho_s(\alpha\leftrightarrow I_g)$ for a few galaxy properties 
$I_g=(m_{\rm gas},\sigma_V,\dot{M}_{\rm{BH}},{\rm SFR})$ extracted directly from the IllustrisTNG simulation data (i.e. they are not post-processed quantities). 
Here again, the correlations are weak ($|\rho_s|$ rarely exceeds 0.2), albeit statistically significant up to $M\sim 10^{12}\ M_\odot/h$ at least. 
The negative $\rho_s$ between $\alpha$ and $\sigma_V$ reflects the fact that $\sigma_V$ measures the depth of the potential well: 
(sub)halos located at the knots of the cosmic web have deeper, more spherically symmetric potential wells and, therefore, larger $\sigma_V$.
As shown in the top panel of Fig.~\ref{fig:env_halo_alpha_bias}, these environments preferentially host halos with low $\alpha$, leading to the observed negative correlation.
By contrast, the positive $\rho_s(\alpha\leftrightarrow {\rm SFR})$ (and $\rho_s(\alpha\leftrightarrow m_\text{gas})$ to a lesser extent) for central galaxies reflects the fact that 
the total gas content and star formation rate are (slightly) higher in environments with larger anisotropy.
Finally, the positive $\rho_s(\alpha\leftrightarrow \dot{M}_{\rm{BH}})$ at halo mass $M\lesssim 10^{11}\hmsun$ explains the (weak) trend between $\alpha$ and 
$L_{\mathrm{H}\alpha}^{\mathrm{AGN}}$ seen in the bottom panel of Fig.~\ref{fig:rho_S_alpha_LHa}. 
A larger $\alpha$ leads to a larger $\dot{M}_{\rm{BH}}$ which, in turn, increases the AGN flux incident on nearby satellite galaxies. 
This enhances their photo-excitation rate and thus raises $L_{\mathrm{H}\alpha}^{\mathrm{AGN}}$.

\begin{figure}
    \centering
    \includegraphics[width=0.42\textwidth]{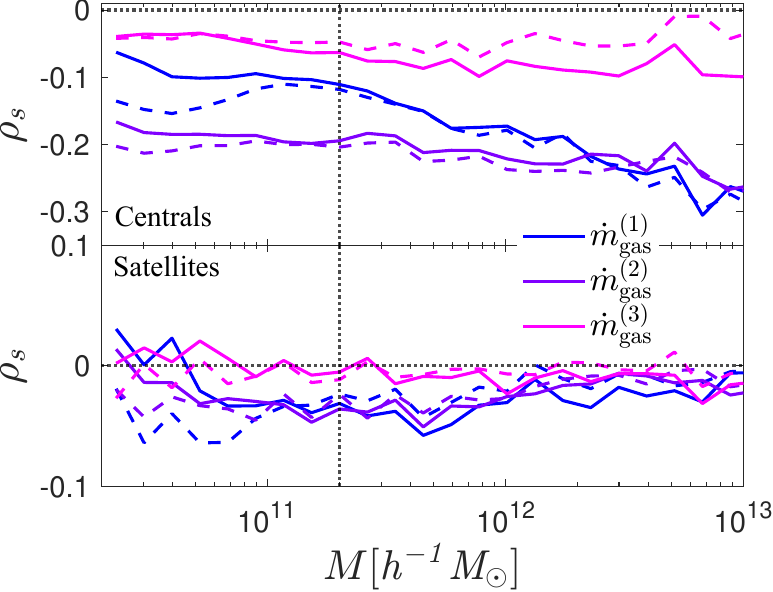}
    \caption{Correlation between $\alpha$ and the gas accretion rate onto individual galaxies for the 3 different settings defined in \ref{sec:galproperties}. The solid lines show correlations with $\alpha_{\rm{V}}$ while dashed lines show the correlations with $\alpha_{\rm{T}}$. The top panel the shows correlation for central galaxies, while the bottom panel shows for satellite galaxies.}
    \label{fig:rho_S_alpha_gas_acc}
\end{figure}
    
To understand this further, Fig.~\ref{fig:rho_S_alpha_gas_acc} shows the correlation coefficient $\rho_s(\alpha \leftrightarrow \dot{m}_{\text{gas}}^{(i)})$ between $\alpha$ and the gas inflow rate computed at 3 different radius (see \S\ref{sec:galproperties} for details).
The correlations are significant for central galaxies solely. In this case, the negative values of $\rho_s$ imply that central galaxies with a higher net gas inflow (i.e. a positive $\dot{ m}_\text{gas}$) are preferentially found in isotropic environments (i.e. low value of $\alpha$) because, in highly anisotropic environments, accretion is restricted to fewer directions. It is also possible that anisotropic environments contain more angular momentum, which inhibits accretion \cite[e.g.][]{song/laigle/etaL:2021}. The signal is strongest when $\dot{ m}_\text{gas}$ is computed at the intermediate radius $4r_{1/2}$. These negative values of $\rho_s$ are, however, not at odds with the positive correlation $\alpha\leftrightarrow {\rm SFR}$ reported in Fig.~\ref{fig:rho_S_alpha_gal_params} because cooling, turbulence and mixing, cloud condensation etc. generally introduce a delay between gas accretion and star formation. 

\begin{figure}
    \centering
    \includegraphics[width=0.5\textwidth]{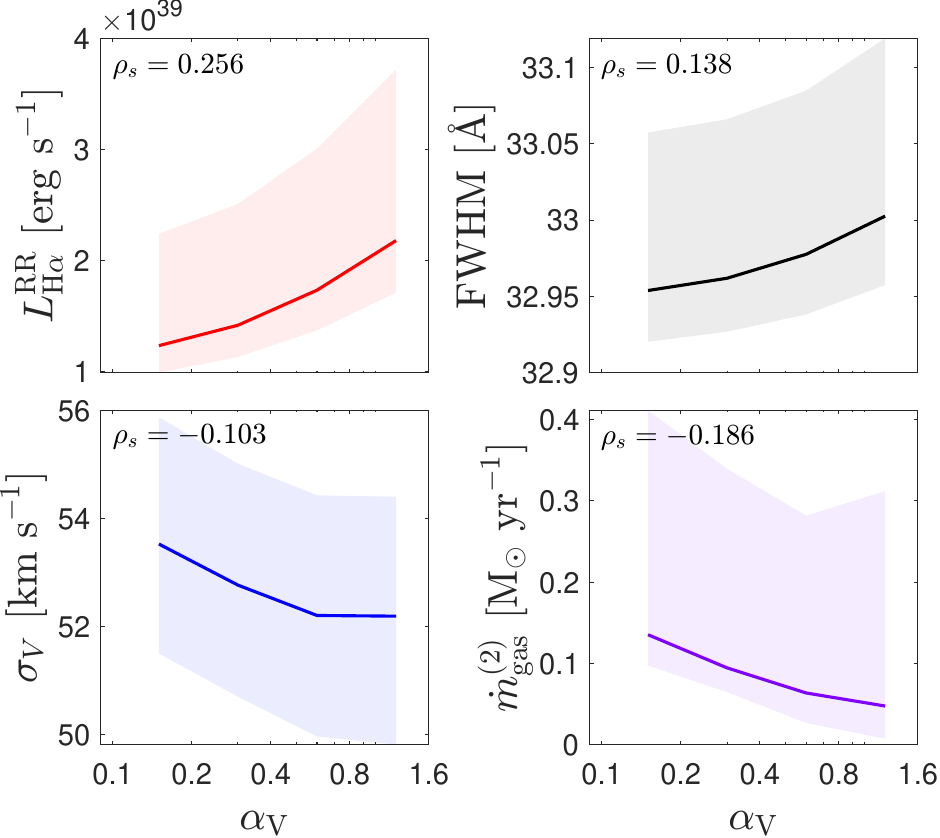}
    \caption{Mean value of galaxy properties computed in bins of $\rm{\alpha_V}$ for central galaxies in the halo mass bin $7\times 10^{10}<M<10^{11} \ M_\odot/h$ (solid line) and the standard deviation of the scatter (shaded). In each panel, the insert quotes the Spearman rank correlation for the scatter.}
    \label{fig:scatter}
\end{figure}

Finally, Fig.~\ref{fig:scatter} displays the mean value of a selected subset of galaxy properties as a function of $\alpha_{\rm T}$ for halos of mass $7\times 10^{10}< M <10^{11}\hmsun$. The shaded area indicates the scatter around the mean on a galaxy-by-galaxy basis. We have also quoted the corresponding values of the Spearman correlation coefficient $\rho_s$. This emphasizes that the correlations reported in this Section are weak. Nonetheless, like e.g. intrinsic alignments in weak lensing data, weak correlations on an object-by-object basis can imprint a detectable signal in the statistics of a large sample.

\subsection{Scale-dependent bias}
\label{sec:scaledependentbias}

\begin{figure}
    \centering
    \includegraphics[width=0.32\textwidth]{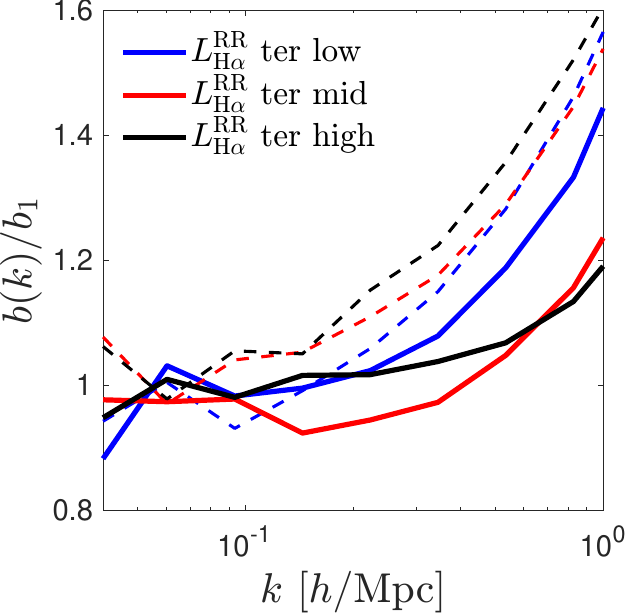}
    \includegraphics[width=0.32\textwidth]{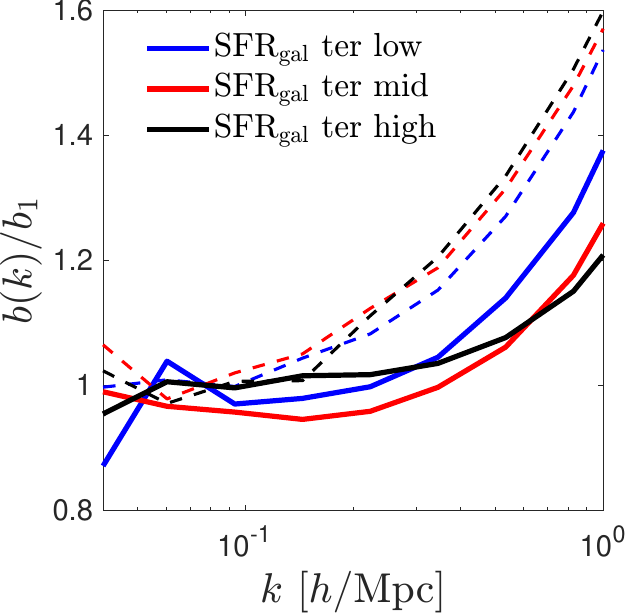}
    \includegraphics[width=0.32\textwidth]{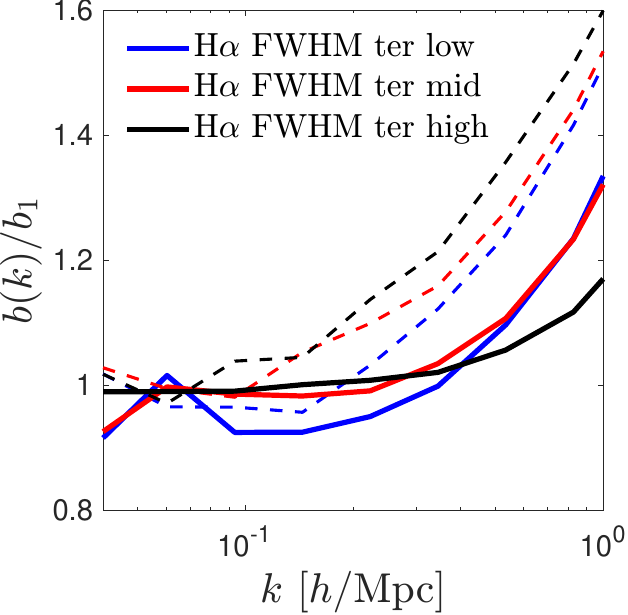}
    \caption{Normalized scale dependent bias of ELGs with $L_{\Halpha} > 10^{41}\ergss $ when split into terciles of $\Halpha$ radiative-recombination luminosity (top), total galactic SFR (middle) and $\Halpha$ line FWHM (bottom). Solid lines show results for central ELGs, and dashed for satellites.}
    \label{fig:ELG_bias_terciles_galaxy_properties}
\end{figure}

To see this, it is instructive to look at the scale-dependence of $b(k)$ as a function of galaxy properties $I_g$. Fig.~\ref{fig:ELG_bias_terciles_galaxy_properties} displays the ratio $b(k)/b_1$ for central (solid curves) and satellite (dashed curves) ELGs selected by $L_{\Halpha}^\text{RR}$, ${\rm SFR}_\text{gal}$ and $\Halpha$ line FWHM (from top to bottom). The mocks have been split into terciles according to the color code shown in the figure. While the satellite galaxies exhibit a strong scale-dependent bias above $k\sim 0.1\hmmpc$, $b(k)$ is weaker for central galaxies. 
Furthermore, central galaxies in the highest tercile of $L_{\Halpha}^\text{RR}$, ${\rm SFR}_\text{gal}$ and $\Halpha$ line FWHM have the weakest scale-dependence, while it is the opposite for satellite galaxies. This scale-dependence reflects the fact that bias parameters beyond the linear bias $b_1$ also correlate with the anisotropy parameter \citep[see e.g.][for measurements of $b_2$ as a function of $\alpha$]{Ramakrishnan_2020}.  Note that, as one transitions from pair counts in separate halos to pairs in the same halo, one expects to see a strong scale-dependent bias. This transition is typically associated with the halo virial radius, for which the corresponding scale is $k\propto 1/r_{\rm vir}$.  This is larger than the $k\sim 0.1\hmmpc$ shown here and, therefore, unlikely to explain our satellite measurements --- unless the relevant scale is the one that sets $\alpha$. 

\section{Conclusions}
\label{sec:conclusions}

In this paper, we have explored the correlation between the properties of $\Halpha$ emitters and their environment using an approach which differs from recent works \citep[e.g.,][]{hadzhiyska/tachella/etal:2021} in two ways: (i) we use a $\Halpha$ emission line model including sources of emission other than HII regions (ii) we quantify the properties of the environment using the anisotropy parameter $\alpha$ defined in \citet{paranjape/etal:2018} rather than the environment density. 
Our results are insensitive to the exact definition of $\alpha$, whether it is computed from the tidal shear or the strain rate tensor. When the normalization of the latter is adjusted such that both definitions are equal in linear theory, they yield similar results for the cosmic web classification. We have also verified that the environmental classifications into knots, filaments, sheets, and voids—constructed from local estimates of $\alpha$ are consistent with those provided in the IllustrisTNG data release, which are based on the DisPerSE algorithm \citep{DisPerSE}.

After performing a few validation tests using dark matter halos and the results from \cite{Ramakrishnan_2019} as a benchmark, we 
apply our approach to mock ELGs extracted from the hydrodynamical suite of simulations IllustrisTNG. For each ELG, we 
assign the value of $\alpha$ obtained for the parent host halo using the halo mass-dependent smoothing radius advocated by \cite{Ramakrishnan_2019}.
We 
find that $>90$\% of the luminous ELGs with $L_{\Halpha}\geq 10^{42}\ergss$ reside in knots (low $\alpha$) and filaments (intermediate $\alpha$), while the rest occupies sheets (high $\alpha$). ELGs with a larger $\alpha$ are more strongly biased on large scales (i.e. their $b_1$ is larger), as expected. 

We have quantified how galaxy properties --- including the different physical sources of $\Halpha$ emission and the $\Halpha$ line FWHM --- correlate with $\alpha$ using Spearman's rank correlation coefficient $\rho_s$. We 
split our results between central and satellite galaxies.
All the correlations are weak ($|\rho_s|<0.5$), but they are statistically significant up to a host halo mass $M=10^{12}\ M_\odot/h$ when measured as a function of $M$. 
They provide clues for the physics of galaxy formation, at least, as it is implemented in IllustrisTNG.
At fixed halo mass, the correlation between RR-induced $\Halpha$ emission from central galaxies and $\alpha$ is the strongest for $M\lesssim 10^{12}\ M_\odot/h$. This correlation, which vanishes nearly entirely when centrals are selected by total $\Halpha$ luminosity, is related to the formation of cosmological-scale accretion shocks in sheets and filaments \citep{Mo2010Galaxy,mandelker/etal:2019,ramsoy/etal:2021,pasha/etal:2023,hasan/etal:2023,hasan/etal:2024}.
For satellite galaxies, $\alpha$ exhibits the strongest correlation with $L_{\Halpha}^\text{AGN}$.
The correlations between galaxy properties and anisotropy parameter are also reflected in the scale-dependence of the ELG bias $b(k)$. For central galaxies with $\Halpha$ luminosity $L_{\Halpha}\geq 10^{41}\ergss$ for example, the scale-dependence of $b(k)$ is weaker for those with high $L_{\Halpha}$, SFR or $\Halpha$ FWHM.

\cite{hadzhiyska/tachella/etal:2021} performed a similar analysis, albeit using standard luminosity-SFR relations \citep[e.g.,][]{Kennicutt_1983} to compute the $\Halpha$ luminosity of mock galaxies in IllustrisTNG-300-1. 
More specifically, they considered the specific star formation rate (sSFR) given by the ratio of SFR to total stellar mass

Furthermore, they adopted a fixed filtering scale of $4\hmpc$ significantly larger than our mass-dependent $\text{R}_{\text{G}}(M)$ --- which varies in the range $0.25\lesssim \rm{R_G}\lesssim 2\hmpc$ for halos in mass range $2\times 10^{11}<M<10^{14}\hmsun$ --- to classify galaxies into knots, filaments, etc. 
Finally, they did not work with the anisotropy parameter $\alpha$ to quantify the correlation between large scale clustering and local environment, but considered instead more traditional proxies such as the environment density. Nonetheless, our conclusions are consistent with theirs wherever there is overlap. 

Would it be possible to select ELGs with $\Halpha$ emission dominated by a specific source (e.g., RR, AGN, etc.) using other spectral signatures? For ELGs dominated by AGN-driven emission, one would also expect significant $\Halpha$ emission from what is referred to as their broad (and narrow) line region BLR (NLR), which our model does not fully take into account. \cite{rapoport/etal:2025} showed the minor effect of excluding AGN (a small fraction) from the calculation. In reality, bright AGNs are obvious, but low luminosity or obscured AGN where the broad lines are not detected unambiguously are harder to distinguish from non-AGN galaxies, and one can think of excluding the central pixel of the galaxy to exclude AGN contributions.

Finally, the strong dependence of $b(k)$ on the environment could prove useful for multi-tracer constraints on cosmology \citep[e.g.,][]{seljak:2009,mcdonald/seljak:2009,white/song/percival:2009,hamaus/etal:2010,hamaus/etal:2011,cai/etal:2011,blake/etal:2013,abramo/leonard:2013,ferramacho/etal:2014,abramo/etal:2016,yamauchi/etal:2017,zhao/etal:2021,schaan/white:2021,sullivan/etal:2023,chen/etal:2024,montero-dorta/rodriguez:2024,EuclidMultiTracer:2024,fang/cai/etal:2024}. To maximize the gains of this approach, the product $(\bar n P)$ should (vastly) exceed unity for each tracer. Using both the $\alpha_{\rm V}$ and $\alpha_{\rm T}$ classifications, we find $\bar n P\gtrsim 10$ (resp. $\bar n P\gtrsim 100$) at $k\sim k_\text{eq}$ for $z=1$ ELGs with luminosity $L_{\Halpha}>10^{42}\ergss$ (resp. $L_{\Halpha}>10^{41}\ergss$) residing in knots or filaments.

Multi-tracer analyses are expected to be particularly useful for constraining the level of primordial non-Gaussianity in the initial conditions \citep{hamaus/etal:2011}.  Previous work on this problem has looked at environment defined by large-scale density only, finding only modest gains in constraining power \citep{densplitPNG}.  This is thought to be because the tracers populate only a narrow range of halo masses (the most massive halos).  Our results provide strong motivation for using ELGs to constrain primordial non-Gaussianity because we have shown that (a) $\Halpha$ emitters span a wide range of halo masses and (b) they present a large environmental dependence that is not confined to density alone.

\section{Acknowledgements}

We thank Shmuel Bialy, Nir Mandelker, Adi Nusser, Aseem Paranjape and Sujatha Ramakrishnan for helpful discussions and comments on the manuscript. The Technion team 
acknowledges support from the Israel Science Foundation (grant nos. 2562/20 and 2617/25).

\appendix

\section{Full Width Half Maximum}
\label{appendix:FWHM}

To generate synthetic $\Halpha$ line profiles, we sum contributions from individual cells in a galaxy using the spatially-resolved emission model described in \cite{rapoport/etal:2025}. First, we compute an intrinsic spectrum upon assigning to each cell a thermal profile
\begin{equation}
    \phi_\text{i}(\nu) = \sqrt{\frac{m_pc^2}{2kT_\text{i}\nu_{\text{cell,i}}^2}}e^{-\frac{m_pc^2(\nu-\nu_{\text{cell,i}})^2}{2kT_\text{i}\nu_{\text{cell,i}}^2}} \;,
\end{equation}
where $m_p$ is the proton mass, $c$ the speed of light, and $T_\text{i}$ the cell temperature. $\nu_{\text{cell,i}}$ is the $\Halpha$ central frequency in that cell, which is computed from the cell's redshift $z_{\text{cell,i}}$ (with corrections for large galaxies, using the redshift-distance relation) and the peculiar velocity component $v_{\text{pec,i}}$ along the observer's line of sight, 
according to
\begin{equation}
    \nu_{\text{cell,i}} = \nu_{\Halpha}\frac{(1-v_{\text{pec,i}}/c)}{1+z_{\text{cell,i}}} \;.
\end{equation}
We adopt a fixed temperature $T_\text{i}=10^4\Kel$ for star forming cells. The luminosity density of the cell (in units of $\rm{erg\ s^{-1}\ \AA^{-1}}$) is thus given by
\begin{equation}
    L_{\lambda ,\Halpha}^i=\frac{1}{4\pi}\frac{c}{\lambda^2}\phi_i(\nu\equiv c/\lambda) L_{\Halpha}^i \;,
\end{equation}
where $L_{\Halpha}^\text{i}$ is the $\Halpha$ luminosity in the cell. The intrinsic spectrum  of a galaxy is the sum of all contributing cells $L_{\lambda ,\Halpha} = \sum_{\text{i}} L_{\lambda ,\Halpha}^\text{i}$. The sampling wavelengths $\lambda$ are constructed in an adaptive manner such that no contributions are missed, and the sampling resolution is fixed to $1 \rm{\AA}$. Finally, the intrinsic spectrum is smoothed using a Gaussian filter with a spectral resolution $\lambda/\Delta \lambda = 400$ matching the resolving power of the \textit{Euclid} NISP spectrometer~\citep{euclidcollaboration2024,EuclidNISP}. Due to this moderate spectral resolution, the resulting smoothed line profiles are mostly featureless except for a single peak with a well-defined FWHM.

\section{Numerical convergence}
\label{appendix:TNG50}

\begin{figure}
    \centering
    \includegraphics[width=0.4\textwidth]{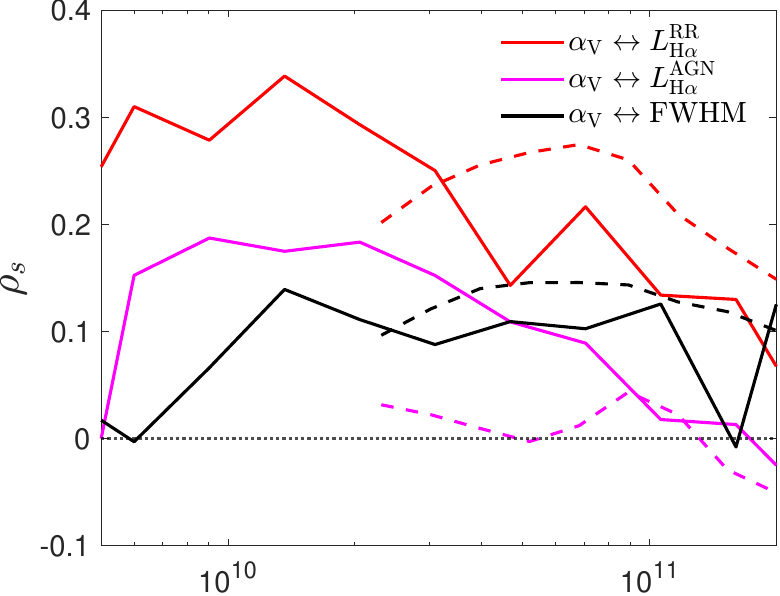}
    \caption{Spearman correlation coefficient $\rho_s$ for the correlation between $\alpha_{\rm{V}}$ and the RR-, AGN-induced $\Halpha$ luminosities as well as the $\Halpha$ line FWHM for central galaxies. Results are shown for TNG50-1 (solid curves) and TNG300-1 (dashed curves).}
    \label{fig:alpha_LHa_TNG50}
\end{figure}

To asses the impact of numerical resolution, we have computed several of the correlation coefficients reported in Fig.~\ref{fig:rho_S_alpha_LHa} using the TNG50-1 simulation, which has a volume $\sim 200$ times smaller and, therefore, allows us to resolve $\rm{R_G}(M)$ down to a halo mass $M\sim 10^{10}\hmsun$ with $\gtrsim 10$ cells using a cubical mesh of size $800^3$. Results are shown in Fig.~\ref{fig:alpha_LHa_TNG50} for the correlation between $\alpha_{\rm{V}}$ with the RR- and AGN-induced $\Halpha$ luminosity as well as the $\Halpha$ line FWHM. Note that we have considered central galaxies solely since satellites are rare in halos of mass $M\lesssim 10^{11}\hmsun$. Although the correlation coefficients $\rho_s$ computed from TNG50-1 are noisier due to the much smaller volume available, they are consistent with the measurements extracted from TNG300-1 wherever there is overlap. 

\bibliography{references}

@article{Ramakrishnan_2019,
   title={Cosmic web anisotropy is the primary indicator of halo assembly bias},
   volume={489},
   ISSN={1365-2966},
   url={http://dx.doi.org/10.1093/mnras/stz2344},
   DOI={10.1093/mnras/stz2344},
   number={3},
   journal={Monthly Notices of the Royal Astronomical Society},
   publisher={Oxford University Press (OUP)},
   author={Ramakrishnan, Sujatha and Paranjape, Aseem and Hahn, Oliver and Sheth, Ravi K},
   year={2019},
   month=aug, pages={2977–2996} }

@ARTICLE{Ramakrishnan_2020,
       author = {{Ramakrishnan}, Sujatha and {Paranjape}, Aseem},
        title = "{Separate Universe calibration of the dependence of halo bias on cosmic web anisotropy}",
      journal = {\mnras},
         year = 2020,
        month = dec,
       volume = {499},
       number = {3},
        pages = {4418-4431},
          doi = {10.1093/mnras/staa2999},
archivePrefix = {arXiv},
       eprint = {2007.03711},
 primaryClass = {astro-ph.CO},
       adsurl = {https://ui.adsabs.harvard.edu/abs/2020MNRAS.499.4418R}
}

@article{Nelson_2019,
   title={First results from the TNG50 simulation: galactic outflows driven by supernovae and black hole feedback},
   volume={490},
   ISSN={1365-2966},
   url={http://dx.doi.org/10.1093/mnras/stz2306},
   DOI={10.1093/mnras/stz2306},
   number={3},
   journal={Monthly Notices of the Royal Astronomical Society},
   publisher={Oxford University Press (OUP)},
   author={Nelson, Dylan and Pillepich, Annalisa and Springel, Volker and Pakmor, Rüdiger and Weinberger, Rainer and Genel, Shy and Torrey, Paul and Vogelsberger, Mark and Marinacci, Federico and Hernquist, Lars},
   year={2019},
   month=aug, pages={3234–3261} }

@article{DisPerSE,
    author = {Sousbie, T.},
    title = "{The persistent cosmic web and its filamentary structure – I. Theory and implementation}",
    journal = {Monthly Notices of the Royal Astronomical Society},
    volume = {414},
    number = {1},
    pages = {350-383},
    year = {2011},
    month = {06},
    issn = {0035-8711},
    doi = {10.1111/j.1365-2966.2011.18394.x},
    url = {https://doi.org/10.1111/j.1365-2966.2011.18394.x},
    eprint = {https://academic.oup.com/mnras/article-pdf/414/1/350/3809858/mnras0414-0350.pdf},
}

@ARTICLE{Dressler1980,
       author = {{Dressler}, A.},
        title = "{Galaxy morphology in rich clusters: implications for the formation and evolution of galaxies.}",
      journal = {\apj},
         year = 1980,
        month = mar,
       volume = {236},
        pages = {351-365},
          doi = {10.1086/157753},
       adsurl = {https://ui.adsabs.harvard.edu/abs/1980ApJ...236..351D}
}

@article{Hoffman_2012,
   title={A kinematic classification of the cosmic web: The cosmic web},
   volume={425},
   ISSN={0035-8711},
   url={http://dx.doi.org/10.1111/j.1365-2966.2012.21553.x},
   DOI={10.1111/j.1365-2966.2012.21553.x},
   number={3},
   journal={Monthly Notices of the Royal Astronomical Society},
   publisher={Oxford University Press (OUP)},
   author={Hoffman, Yehuda and Metuki, Ofer and Yepes, Gustavo and Gottlöber, Stefan and Forero-Romero, Jaime E. and Libeskind, Noam I. and Knebe, Alexander},
   year={2012},
   month=aug, pages={2049–2057} }

@article{Libeskind_2012,
   title={The velocity shear tensor: tracer of halo alignment},
   volume={428},
   ISSN={1365-2966},
   url={http://dx.doi.org/10.1093/mnras/sts216},
   DOI={10.1093/mnras/sts216},
   number={3},
   journal={Monthly Notices of the Royal Astronomical Society},
   publisher={Oxford University Press (OUP)},
   author={Libeskind, N. I. and Hoffman, Y. and Forero-Romero, J. and Gottlober, S. and Knebe, A. and Steinmetz, M. and Klypin, A.},
   year={2012},
   month=nov, pages={2489–2499} }

@ARTICLE{hadzhiyska/tachella/etal:2021,
       author = {{Hadzhiyska}, Boryana and {Tacchella}, Sandro and {Bose}, Sownak and {Eisenstein}, Daniel J.},
        title = "{The galaxy-halo connection of emission-line galaxies in IllustrisTNG}",
      journal = {\mnras},
         year = 2021,
        month = apr,
       volume = {502},
       number = {3},
        pages = {3599-3617},
          doi = {10.1093/mnras/stab243},
archivePrefix = {arXiv},
       eprint = {2011.05331},
 primaryClass = {astro-ph.GA},
       adsurl = {https://ui.adsabs.harvard.edu/abs/2021MNRAS.502.3599H}
}

@misc{spergel2015,
      title={Wide-Field InfrarRed Survey Telescope-Astrophysics Focused Telescope Assets WFIRST-AFTA 2015 Report}, 
      author={D. Spergel and N. Gehrels and C. Baltay and D. Bennett and J. Breckinridge and M. Donahue and A. Dressler and B. S. Gaudi and T. Greene and O. Guyon and C. Hirata and J. Kalirai and N. J. Kasdin and B. Macintosh and W. Moos and S. Perlmutter and M. Postman and B. Rauscher and J. Rhodes and Y. Wang and D. Weinberg and D. Benford and M. Hudson and W. -S. Jeong and Y. Mellier and W. Traub and T. Yamada and P. Capak and J. Colbert and D. Masters and M. Penny and D. Savransky and D. Stern and N. Zimmerman and R. Barry and L. Bartusek and K. Carpenter and E. Cheng and D. Content and F. Dekens and R. Demers and K. Grady and C. Jackson and G. Kuan and J. Kruk and M. Melton and B. Nemati and B. Parvin and I. Poberezhskiy and C. Peddie and J. Ruffa and J. K. Wallace and A. Whipple and E. Wollack and F. Zhao},
      year={2015},
      eprint={1503.03757},
      archivePrefix={arXiv},
      primaryClass={astro-ph.IM},
      url={https://arxiv.org/abs/1503.03757}, 
}

@ARTICLE{euclidcollaboration2024,
       author = {{Euclid Collaboration} and {Mellier}, Y. and {Abdurro'uf} and {Acevedo Barroso}, J.~A. and {Ach{\'u}carro}, A. and {Adamek}, J. and {Adam}, R. and {Addison}, G.~E. and {Aghanim}, N. and {Aguena}, M. and {Ajani}, V. and {Akrami}, Y. and {Al-Bahlawan}, A. and {Alavi}, A. and {Albuquerque}, I.~S. and {Alestas}, G. and {Alguero}, G. and {Allaoui}, A. and {Allen}, S.~W. and {Allevato}, V. and {Alonso-Tetilla}, A.~V. and {Altieri}, B. and {Alvarez-Candal}, A. and {Amara}, A. and {Amendola}, L. and {Amiaux}, J. and {Andika}, I.~T. and {Andreon}, S. and {Andrews}, A. and {Angora}, G. and {Angulo}, R.~E. and {Annibali}, F. and {Anselmi}, A. and {Anselmi}, S. and {Arcari}, S. and {Archidiacono}, M. and {Aric{\`o}}, G. and {Arnaud}, M. and {Arnouts}, S. and {Asgari}, M. and {Asorey}, J. and {Atayde}, L. and {Atek}, H. and {Atrio-Barandela}, F. and {Aubert}, M. and {Aubourg}, E. and {Auphan}, T. and {Auricchio}, N. and {Aussel}, B. and {Aussel}, H. and {Avelino}, P.~P. and {Avgoustidis}, A. and {Avila}, S. and {Awan}, S. and {Azzollini}, R. and {Baccigalupi}, C. and {Bachelet}, E. and {Bacon}, D. and {Baes}, M. and {Bagley}, M.~B. and {Bahr-Kalus}, B. and {Balaguera-Antolinez}, A. and {Balbinot}, E. and {Balcells}, M. and {Baldi}, M. and {Baldry}, I. and {Balestra}, A. and {Ballardini}, M. and {Ballester}, O. and {Balogh}, M. and {Ba{\~n}ados}, E. and {Barbier}, R. and {Bardelli}, S. and {Barreiro}, T. and {Barriere}, J. -C. and {Barros}, B.~J. and {Barthelemy}, A. and {Bartolo}, N. and {Basset}, A. and {Battaglia}, P. and {Battisti}, A.~J. and {Baugh}, C.~M. and {Baumont}, L. and {Bazzanini}, L. and {Beaulieu}, J. -P. and {Beckmann}, V. and {Belikov}, A.~N. and {Bel}, J. and {Bellagamba}, F. and {Bella}, M. and {Bellini}, E. and {Benabed}, K. and {Bender}, R. and {Benevento}, G. and {Bennett}, C.~L. and {Benson}, K. and {Bergamini}, P. and {Bermejo-Climent}, J.~R. and {Bernardeau}, F. and {Bertacca}, D. and {Berthe}, M. and {Berthier}, J. and {Bethermin}, M. and {Beutler}, F. and {Bevillon}, C. and {Bhargava}, S. and {Bhatawdekar}, R. and {Bisigello}, L. and {Biviano}, A. and {Blake}, R.~P. and {Blanchard}, A. and {Blazek}, J. and {Blot}, L. and {Bosco}, A. and {Bodendorf}, C. and {Boenke}, T. and {B{\"o}hringer}, H. and {Bolzonella}, M. and {Bonchi}, A. and {Bonici}, M. and {Bonino}, D. and {Bonino}, L. and {Bonvin}, C. and {Bon}, W. and {Booth}, J.~T. and {Borgani}, S. and {Borlaff}, A.~S. and {Borsato}, E. and {Bosco}, A. and {Bose}, B. and {Botticella}, M.~T. and {Boucaud}, A. and {Bouche}, F. and {Boucher}, J.~S. and {Boutigny}, D. and {Bouvard}, T. and {Bouy}, H. and {Bowler}, R.~A.~A. and {Bozza}, V. and {Bozzo}, E. and {Branchini}, E. and {Brau-Nogue}, S. and {Brekke}, P. and {Bremer}, M.~N. and {Brescia}, M. and {Breton}, M. -A. and {Brinchmann}, J. and {Brinckmann}, T. and {Brockley-Blatt}, C. and {Brodwin}, M. and {Brouard}, L. and {Brown}, M.~L. and {Bruton}, S. and {Bucko}, J. and {Buddelmeijer}, H. and {Buenadicha}, G. and {Buitrago}, F. and {Burger}, P. and {Burigana}, C. and {Busillo}, V. and {Busonero}, D. and {Cabanac}, R. and {Cabayol-Garcia}, L. and {Cagliari}, M.~S. and {Caillat}, A. and {Caillat}, L. and {Calabrese}, M. and {Calabro}, A. and {Calderone}, G. and {Calura}, F. and {Camacho Quevedo}, B. and {Camera}, S. and {Campos}, L. and {Canas-Herrera}, G. and {Candini}, G.~P. and {Cantiello}, M. and {Capobianco}, V. and {Cappellaro}, E. and {Cappelluti}, N. and {Cappi}, A. and {Caputi}, K.~I. and {Cara}, C. and {Carbone}, C. and {Cardone}, V.~F. and {Carella}, E. and {Carlberg}, R.~G. and {Carle}, M. and {Carminati}, L. and {Caro}, F. and {Carrasco}, J.~M. and {Carretero}, J. and {Carrilho}, P. and {Carron Duque}, J. and {Carry}, B. and {Carvalho}, A. and {Carvalho}, C.~S. and {Casas}, R. and {Casas}, S. and {Casenove}, P. and {Casey}, C.~M. and {Cassata}, P. and {Castander}, F.~J. and {Castelao}, D. and {Castellano}, M. and {Castiblanco}, L. and {Castignani}, G. and {Castro}, T. and {Cavet}, C. and {Cavuoti}, S. and {Chabaud}, P. -Y. and {Chambers}, K.~C. and {Charles}, Y. and {Charlot}, S. and {Chartab}, N. and {Chary}, R. and {Chaumeil}, F. and {Cho}, H. and {Chon}, G. and {Ciancetta}, E. and {Ciliegi}, P. and {Cimatti}, A. and {Cimino}, M. and {Cioni}, M. -R.~L. and {Claydon}, R. and {Cleland}, C. and {Cl{\'e}ment}, B. and {Clements}, D.~L. and {Clerc}, N. and {Clesse}, S. and {Codis}, S. and {Cogato}, F. and {Colbert}, J. and {Cole}, R.~E. and {Coles}, P. and {Collett}, T.~E. and {Collins}, R.~S. and {Colodro-Conde}, C. and {Colombo}, C. and {Combes}, F. and {Conforti}, V. and {Congedo}, G. and {Conseil}, S. and {Conselice}, C.~J. and {Contarini}, S. and {Contini}, T. and {Conversi}, L. and {Cooray}, A.~R. and {Copin}, Y. and {Corasaniti}, P. -S. and {Corcho-Caballero}, P. and {Corcione}, L. and {Cordes}, O. and {Corpace}, O. and {Correnti}, M. and {Costanzi}, M. and {Costille}, A. and {Courbin}, F. and {Courcoult Mifsud}, L. and {Courtois}, H.~M. and {Cousinou}, M. -C. and {Covone}, G. and {Cowell}, T. and {Cragg}, C. and {Cresci}, G. and {Cristiani}, S. and {Crocce}, M. and {Cropper}, M. and {E Crouzet}, P. and {Csizi}, B. and {Cuby}, J. -G. and {Cucchetti}, E. and {Cucciati}, O. and {Cuillandre}, J. -C. and {Cunha}, P.~A.~C. and {Cuozzo}, V. and {Daddi}, E. and {D'Addona}, M. and {Dafonte}, C. and {Dagoneau}, N. and {Dalessandro}, E. and {Dalton}, G.~B. and {D'Amico}, G. and {Dannerbauer}, H. and {Danto}, P. and {Das}, I. and {Da Silva}, A. and {da Silva}, R. and {Daste}, G. and {Davies}, J.~E. and {Davini}, S. and {de Boer}, T. and {Decarli}, R. and {De Caro}, B. and {Degaudenzi}, H. and {Degni}, G. and {de Jong}, J.~T.~A. and {de la Bella}, L.~F. and {de la Torre}, S. and {Delhaise}, F. and {Delley}, D. and {Delucchi}, G. and {De Lucia}, G. and {Denniston}, J. and {De Paolis}, F. and {De Petris}, M. and {Derosa}, A. and {Desai}, S. and {Desjacques}, V. and {Despali}, G. and {Desprez}, G. and {De Vicente-Albendea}, J. and {Deville}, Y. and {Dias}, J.~D.~F. and {D{\'\i}az-S{\'a}nchez}, A. and {Diaz}, J.~J. and {Di Domizio}, S. and {Diego}, J.~M. and {Di Ferdinando}, D. and {Di Giorgio}, A.~M. and {Dimauro}, P. and {Dinis}, J. and {Dolag}, K. and {Dolding}, C. and {Dole}, H. and {Dom{\'\i}nguez S{\'a}nchez}, H. and {Dor{\'e}}, O. and {Dournac}, F. and {Douspis}, M. and {Dreihahn}, H. and {Droge}, B. and {Dryer}, B. and {Dubath}, F. and {Duc}, P. -A. and {Ducret}, F. and {Duffy}, C. and {Dufresne}, F. and {Duncan}, C.~A.~J. and {Dupac}, X. and {Duret}, V. and {Durrer}, R. and {Durret}, F. and {Dusini}, S. and {Ealet}, A. and {Eggemeier}, A. and {Eisenhardt}, P.~R.~M. and {Elbaz}, D. and {Elkhashab}, M.~Y. and {Ellien}, A. and {Endicott}, J. and {Enia}, A. and {Erben}, T. and {Escartin Vigo}, J.~A. and {Escoffier}, S. and {Escudero Sanz}, I. and {Essert}, J. and {Ettori}, S. and {Ezziati}, M. and {Fabbian}, G. and {Fabricius}, M. and {Fang}, Y. and {Farina}, A. and {Farina}, M. and {Farinelli}, R. and {Farrens}, S. and {Faustini}, F. and {Feltre}, A. and {Ferguson}, A.~M.~N. and {Ferrando}, P. and {Ferrari}, A.~G. and {Ferr{\'e}-Mateu}, A. and {Ferreira}, P.~G. and {Ferreras}, I. and {Ferrero}, I. and {Ferriol}, S. and {Ferruit}, P. and {Filleul}, D. and {Finelli}, F. and {Finkelstein}, S.~L. and {Finoguenov}, A. and {Fiorini}, B. and {Flentge}, F. and {Focardi}, P. and {Fonseca}, J. and {Fontana}, A. and {Fontanot}, F. and {Fornari}, F. and {Fosalba}, P. and {Fossati}, M. and {Fotopoulou}, S. and {Fouchez}, D. and {Fourmanoit}, N. and {Frailis}, M. and {Fraix-Burnet}, D. and {Franceschi}, E. and {Franco}, A. and {Franzetti}, P. and {Freihoefer}, J. and {Frittoli}, G. and {Frugier}, P. -A. and {Frusciante}, N. and {Fumagalli}, A. and {Fumagalli}, M. and {Fumana}, M. and {Fu}, Y. and {Gabarra}, L. and {Galeotta}, S. and {Galluccio}, L. and {Ganga}, K. and {Gao}, H. and {Garc{\'\i}a-Bellido}, J. and {Garcia}, K. and {Gardner}, J.~P. and {Garilli}, B. and {Gaspar-Venancio}, L. -M. and {Gasparetto}, T. and {Gautard}, V. and {Gavazzi}, R. and {Gaztanaga}, E. and {Genolet}, L. and {Genova Santos}, R. and {Gentile}, F. and {George}, K. and {Ghaffari}, Z. and {Giacomini}, F. and {Gianotti}, F. and {Gibb}, G.~P.~S. and {Gillard}, W. and {Gillis}, B. and {Ginolfi}, M. and {Giocoli}, C. and {Girardi}, M. and {Giri}, S.~K. and {Goh}, L.~W.~K. and {G{\'o}mez-Alvarez}, P. and {Gonzalez}, A.~H. and {Gonzalez}, E.~J. and {Gonzalez}, J.~C. and {Gouyou Beauchamps}, S. and {Gozaliasl}, G. and {Gracia-Carpio}, J. and {Grandis}, S. and {Granett}, B.~R. and {Granvik}, M. and {Grazian}, A. and {Gregorio}, A. and {Grenet}, C. and {Grillo}, C. and {Grupp}, F. and {Gruppioni}, C. and {Gruppuso}, A. and {Guerbuez}, C. and {Guerrini}, S. and {Guidi}, M. and {Guillard}, P. and {Gutierrez}, C.~M. and {Guttridge}, P. and {Guzzo}, L. and {Gwyn}, S. and {Haapala}, J. and {Haase}, J. and {Haddow}, C.~R. and {Hailey}, M. and {Hall}, A. and {Hall}, D. and {Hamaus}, N. and {Haridasu}, B.~S. and {Harnois-D{\'e}raps}, J. and {Harper}, C. and {Hartley}, W.~G. and {Hasinger}, G. and {Hassani}, F. and {Hatch}, N.~A. and {Haugan}, S.~V.~H. and {H{\"a}u{\ss}ler}, B. and {Heavens}, A. and {Heisenberg}, L. and {Helmi}, A. and {Helou}, G. and {Hemmati}, S. and {Henares}, K. and {Herent}, O. and {Hern{\'a}ndez-Monteagudo}, C. and {Heuberger}, T. and {Hewett}, P.~C. and {Heydenreich}, S. and {Hildebrandt}, H. and {Hirschmann}, M. and {Hjorth}, J. and {Hoar}, J. and {Hoekstra}, H. and {Holland}, A.~D. and {Holliman}, M.~S. and {Holmes}, W. and {Hook}, I. and {Horeau}, B. and {Hormuth}, F. and {Hornstrup}, A. and {Hosseini}, S. and {Hu}, D. and {Hudelot}, P. and {Hudson}, M.~J. and {Huertas-Company}, M. and {Huff}, E.~M. and {Hughes}, A.~C.~N. and {Humphrey}, A. and {Hunt}, L.~K. and {Huynh}, D.~D. and {Ibata}, R. and {Ichikawa}, K. and {Iglesias-Groth}, S. and {Ilbert}, O. and {Ili{\'c}}, S. and {Ingoglia}, L. and {Iodice}, E. and {Israel}, H. and {Israelsson}, U.~E. and {Izzo}, L. and {Jablonka}, P. and {Jackson}, N. and {Jacobson}, J. and {Jafariyazani}, M. and {Jahnke}, K. and {Jansen}, H. and {Jarvis}, M.~J. and {Jasche}, J. and {Jauzac}, M. and {Jeffrey}, N. and {Jhabvala}, M. and {Jimenez-Teja}, Y. and {Jimenez Mu{\~n}oz}, A. and {Joachimi}, B. and {Johansson}, P.~H. and {Joudaki}, S. and {Jullo}, E. and {Kajava}, J.~J.~E. and {Kang}, Y. and {Kannawadi}, A. and {Kansal}, V. and {Karagiannis}, D. and {K{\"a}rcher}, M. and {Kashlinsky}, A. and {Kazandjian}, M.~V. and {Keck}, F. and {Keih{\"a}nen}, E. and {Kerins}, E. and {Kermiche}, S. and {Khalil}, A. and {Kiessling}, A. and {Kiiveri}, K. and {Kilbinger}, M. and {Kim}, J. and {King}, R. and {Kirkpatrick}, C.~C. and {Kitching}, T. and {Kluge}, M. and {Knabenhans}, M. and {Knapen}, J.~H. and {Knebe}, A. and {Kneib}, J. -P. and {Kohley}, R. and {Koopmans}, L.~V.~E. and {Koskinen}, H. and {Koulouridis}, E. and {Kou}, R. and {Kov{\'a}cs}, A. and {Kova\{{\v{c}}\}i{\'c}}, I. and {Kowalczyk}, A. and {Koyama}, K. and {Kraljic}, K. and {Krause}, O. and {Kruk}, S. and {Kubik}, B. and {Kuchner}, U. and {Kuijken}, K. and {K{\"u}mmel}, M. and {Kunz}, M. and {Kurki-Suonio}, H. and {Lacasa}, F. and {Lacey}, C.~G. and {La Franca}, F. and {Lagarde}, N. and {Lahav}, O. and {Laigle}, C. and {La Marca}, A. and {La Marle}, O. and {Lamine}, B. and {Lam}, M.~C. and {Lan{\c{c}}on}, A. and {Landt}, H. and {Langer}, M. and {Lapi}, A. and {Larcheveque}, C. and {Larsen}, S.~S. and {Lattanzi}, M. and {Laudisio}, F. and {Laugier}, D. and {Laureijs}, R. and {Lavaux}, G. and {Lawrenson}, A. and {Lazanu}, A. and {Lazeyras}, T. and {Le Boulc'h}, Q. and {Le Brun}, A.~M.~C. and {Le Brun}, V. and {Leclercq}, F. and {Lee}, S. and {Le Graet}, J. and {Legrand}, L. and {Leirvik}, K.~N. and {Le Jeune}, M. and {Lembo}, M. and {Le Mignant}, D. and {Lepinzan}, M.~D. and {Lepori}, F. and {Lesci}, G.~F. and {Lesgourgues}, J. and {Leuzzi}, L. and {Levi}, M.~E. and {Liaudat}, T.~I. and {Libet}, G. and {Liebing}, P. and {Ligori}, S. and {Lilje}, P.~B. and {Lin}, C. -C. and {Linde}, D. and {Linder}, E. and {Lindholm}, V. and {Linke}, L. and {Li}, S. -S. and {Liu}, S.~J. and {Lloro}, I. and {Lobo}, F.~S.~N. and {Lodieu}, N. and {Lombardi}, M. and {Lombriser}, L. and {Lonare}, P. and {Longo}, G. and {L{\'o}pez-Caniego}, M. and {Lopez Lopez}, X. and {Alvarez}, J. Lorenzo and {Loureiro}, A. and {Loveday}, J. and {Lusso}, E. and {Macias-Perez}, J. and {Maciaszek}, T. and {Magliocchetti}, M. and {Magnard}, F. and {Magnier}, E.~A. and {Magro}, A. and {Mahler}, G. and {Mainetti}, G. and {Maino}, D. and {Maiorano}, E. and {Maiorano}, E. and {Malavasi}, N. and {Mamon}, G.~A. and {Mancini}, C. and {Mandelbaum}, R. and {Manera}, M. and {Manj{\'o}n-Garc{\'\i}a}, A. and {Mannucci}, F. and {Mansutti}, O. and {Manteiga Outeiro}, M. and {Maoli}, R. and {Maraston}, C. and {Marcin}, S. and {Marcos-Arenal}, P. and {Margalef-Bentabol}, B. and {Marggraf}, O. and {Marinucci}, D. and {Marinucci}, M. and {Markovic}, K. and {Marleau}, F.~R. and {Marpaud}, J. and {Martignac}, J. and {Mart{\'\i}n-Fleitas}, J. and {Martin-Moruno}, P. and {Martin}, E.~L. and {Martinelli}, M. and {Martinet}, N. and {Martin}, H. and {Martins}, C.~J.~A.~P. and {Marulli}, F. and {Massari}, D. and {Massey}, R. and {Masters}, D.~C. and {Matarrese}, S. and {Matsuoka}, Y. and {Matthew}, S. and {Maughan}, B.~J. and {Mauri}, N. and {Maurin}, L. and {Maurogordato}, S. and {McCarthy}, K. and {McConnachie}, A.~W. and {McCracken}, H.~J. and {McDonald}, I. and {McEwen}, J.~D. and {McPartland}, C.~J.~R. and {Medinaceli}, E. and {Mehta}, V. and {Mei}, S. and {Melchior}, M. and {Melin}, J. -B. and {M{\'e}nard}, B. and {Mendes}, J. and {Mendez-Abreu}, J. and {Meneghetti}, M. and {Mercurio}, A. and {Merlin}, E. and {Metcalf}, R.~B. and {Meylan}, G. and {Migliaccio}, M. and {Mignoli}, M. and {Miller}, L. and {Miluzio}, M. and {Milvang-Jensen}, B. and {Mimoso}, J.~P. and {Miquel}, R. and {Miyatake}, H. and {Mobasher}, B. and {Mohr}, J.~J. and {Monaco}, P. and {Mongui{\'o}}, M. and {Montoro}, A. and {Mora}, A. and {Moradinezhad Dizgah}, A. and {Moresco}, M. and {Moretti}, C. and {Morgante}, G. and {Morisset}, N. and {Moriya}, T.~J. and {Morris}, P.~W. and {Mortlock}, D.~J. and {Moscardini}, L. and {Mota}, D.~F. and {Moustakas}, L.~A. and {Moutard}, T. and {M{\"u}ller}, T. and {Munari}, E. and {Murphree}, G. and {Murray}, C. and {Murray}, N. and {Musi}, P. and {Nadathur}, S. and {Nagam}, B.~C. and {Nagao}, T. and {Naidoo}, K. and {Nakajima}, R. and {Nally}, C. and {Natoli}, P. and {Navarro-Alsina}, A. and {Navarro Girones}, D. and {Neissner}, C. and {Nersesian}, A. and {Nesseris}, S. and {Nguyen-Kim}, H.~N. and {Nicastro}, L. and {Nichol}, R.~C. and {Nielbock}, M. and {Niemi}, S. -M. and {Nieto}, S. and {Nilsson}, K. and {Noller}, J. and {Norberg}, P. and {Nourizonoz}, A. and {Ntelis}, P. and {Nucita}, A.~A. and {Nugent}, P. and {Nunes}, N.~J. and {Nutma}, T. and {Ocampo}, I. and {Odier}, J. and {Oesch}, P.~A. and {Oguri}, M. and {Magalhaes Oliveira}, D. and {Onoue}, M. and {Oosterbroek}, T. and {Oppizzi}, F. and {Ordenovic}, C. and {Osato}, K. and {Pacaud}, F. and {Pace}, F. and {Padilla}, C. and {Paech}, K. and {Pagano}, L. and {Page}, M.~J. and {Palazzi}, E. and {Paltani}, S. and {Pamuk}, S. and {Pandolfi}, S. and {Paoletti}, D. and {Paolillo}, M. and {Papaderos}, P. and {Pardede}, K. and {Parimbelli}, G. and {Parmar}, A. and {Partmann}, C. and {Pasian}, F. and {Passalacqua}, F. and {Paterson}, K. and {Patrizii}, L. and {Pattison}, C. and {Paulino-Afonso}, A. and {Paviot}, R. and {Peacock}, J.~A. and {Pearce}, F.~R. and {Pedersen}, K. and {Peel}, A. and {Peletier}, R.~F. and {Pellejero Ibanez}, M. and {Pello}, R. and {Penny}, M.~T. and {Percival}, W.~J. and {Perez-Garrido}, A. and {Perotto}, L. and {Pettorino}, V. and {Pezzotta}, A. and {Pezzuto}, S. and {Philippon}, A. and {Piersanti}, O. and {Pietroni}, M. and {Piga}, L. and {Pilo}, L. and {Pires}, S. and {Pisani}, A. and {Pizzella}, A. and {Pizzuti}, L. and {Plana}, C. and {Polenta}, G. and {Pollack}, J.~E. and {Poncet}, M. and {P{\"o}ntinen}, M. and {Pool}, P. and {Popa}, L.~A. and {Popa}, V. and {Popp}, J. and {Porciani}, C. and {Porth}, L. and {Potter}, D. and {Poulain}, M. and {Pourtsidou}, A. and {Pozzetti}, L. and {Prandoni}, I. and {Pratt}, G.~W. and {Prezelus}, S. and {Prieto}, E. and {Pugno}, A. and {Quai}, S. and {Quilley}, L. and {Racca}, G.~D. and {Raccanelli}, A. and {R{\'a}cz}, G. and {Radinovi{\'c}}, S. and {Radovich}, M. and {Ragagnin}, A. and {Ragnit}, U. and {Raison}, F. and {Ramos-Chernenko}, N. and {Ranc}, C. and {Raylet}, N. and {Rebolo}, R. and {Refregier}, A. and {Reimberg}, P. and {Reiprich}, T.~H. and {Renk}, F. and {Renzi}, A. and {Retre}, J. and {Revaz}, Y. and {Reyl{\'e}}, C. and {Reynolds}, L. and {Rhodes}, J. and {Ricci}, F. and {Ricci}, M. and {Riccio}, G. and {Ricken}, S.~O. and {Rissanen}, S. and {Risso}, I. and {Rix}, H. -W. and {Robin}, A.~C. and {Rocca-Volmerange}, B. and {Rocci}, P. -F. and {Rodenhuis}, M. and {Rodighiero}, G. and {Rodriguez Monroy}, M. and {Rollins}, R.~P. and {Romanello}, M. and {Roman}, J. and {Romelli}, E. and {Romero-Gomez}, M. and {Roncarelli}, M. and {Rosati}, P. and {Rosset}, C. and {Rossetti}, E. and {Roster}, W. and {Rottgering}, H.~J.~A. and {Rozas-Fern{\'a}ndez}, A. and {Ruane}, K. and {Rubino-Martin}, J.~A. and {Rudolph}, A. and {Ruppin}, F. and {Rusholme}, B. and {Sacquegna}, S. and {S{\'a}ez-Casares}, I. and {Saga}, S. and {Saglia}, R. and {Sahl{\'e}n}, M. and {Saifollahi}, T. and {Sakr}, Z. and {Salvalaggio}, J. and {Salvaterra}, R. and {Salvati}, L. and {Salvato}, M. and {Salvignol}, J. -C. and {S{\'a}nchez}, A.~G. and {Sanchez}, E. and {Sanders}, D.~B. and {Sapone}, D. and {Saponara}, M. and {Sarpa}, E. and {Sarron}, F. and {Sartori}, S. and {Sassolas}, B. and {Sauniere}, L. and {Sauvage}, M. and {Sawicki}, M. and {Scaramella}, R. and {Scarlata}, C. and {Scharr{\'e}}, L. and {Schaye}, J. and {Schewtschenko}, J.~A. and {Schindler}, J. -T. and {Schinnerer}, E. and {Schirmer}, M. and {Schmidt}, F. and {Schmidt}, F. and {Schmidt}, M. and {Schneider}, A. and {Schneider}, M. and {Schneider}, P. and {Sch{\"o}neberg}, N. and {Schrabback}, T. and {Schultheis}, M. and {Schulz}, S. and {Schwartz}, J. and {Sciotti}, D. and {Scodeggio}, M. and {Scognamiglio}, D. and {Scott}, D. and {Scottez}, V. and {Secroun}, A. and {Sefusatti}, E. and {Seidel}, G. and {Seiffert}, M. and {Sellentin}, E. and {Selwood}, M. and {Semboloni}, E. and {Sereno}, M. and {Serjeant}, S. and {Serrano}, S. and {Shankar}, F. and {Sharples}, R.~M. and {Short}, A. and {Shulevski}, A. and {Shuntov}, M. and {Sias}, M. and {Sikkema}, G. and {Silvestri}, A. and {Simon}, P. and {Sirignano}, C. and {Sirri}, G. and {Skottfelt}, J. and {Slezak}, E. and {Sluse}, D. and {Smith}, G.~P. and {Smith}, L.~C. and {Smith}, R.~E. and {Smit}, S.~J.~A. and {Soldano}, F. and {Solheim}, B.~G.~B. and {Sorce}, J.~G. and {Sorrenti}, F. and {Soubrie}, E. and {Spinoglio}, L. and {Spurio Mancini}, A. and {Stadel}, J. and {Stagnaro}, L. and {Stanco}, L. and {Stanford}, S.~A. and {Starck}, J. -L. and {Stassi}, P. and {Steinwagner}, J. and {Stern}, D. and {Stone}, C. and {Strada}, P. and {Strafella}, F. and {Stramaccioni}, D. and {Surace}, C. and {Sureau}, F. and {Suyu}, S.~H. and {Swindells}, I. and {Szafraniec}, M. and {Szapudi}, I. and {Taamoli}, S. and {Talia}, M. and {Tallada-Cresp{\'\i}}, P. and {Tanidis}, K. and {Tao}, C. and {Tarr{\'\i}o}, P. and {Tavagnacco}, D. and {Taylor}, A.~N. and {Taylor}, J.~E. and {Taylor}, P.~L. and {Teixeira}, E.~M. and {Tenti}, M. and {Teodoro Idiago}, P. and {Teplitz}, H.~I. and {Tereno}, I. and {Tessore}, N. and {Testa}, V. and {Testera}, G. and {Tewes}, M. and {Teyssier}, R. and {Theret}, N. and {Thizy}, C. and {Thomas}, P.~D. and {Toba}, Y. and {Toft}, S. and {Toledo-Moreo}, R. and {Tolstoy}, E. and {Tommasi}, E. and {Torbaniuk}, O. and {Torradeflot}, F. and {Tortora}, C. and {Tosi}, S. and {Tosti}, S. and {Trifoglio}, M. and {Troja}, A. and {Trombetti}, T. and {Tronconi}, A. and {Tsedrik}, M. and {Tsyganov}, A. and {Tucci}, M. and {Tutusaus}, I. and {Uhlemann}, C. and {Ulivi}, L. and {Urbano}, M. and {Vacher}, L. and {Vaillon}, L. and {Valdes}, I. and {Valentijn}, E.~A. and {Valenziano}, L. and {Valieri}, C. and {Valiviita}, J. and {Van den Broeck}, M. and {Vassallo}, T. and {Vavrek}, R. and {Venemans}, B. and {Venhola}, A. and {Ventura}, S. and {Verdoes Kleijn}, G. and {Vergani}, D. and {Verma}, A. and {Vernizzi}, F. and {Veropalumbo}, A. and {Verza}, G. and {Vescovi}, C. and {Vibert}, D. and {Viel}, M. and {Vielzeuf}, P. and {Viglione}, C. and {Viitanen}, A. and {Villaescusa-Navarro}, F. and {Vinciguerra}, S. and {Visticot}, F. and {Voggel}, K. and {von Wietersheim-Kramsta}, M. and {Vriend}, W.~J. and {Wachter}, S. and {Walmsley}, M. and {Walth}, G. and {Walton}, D.~M. and {Walton}, N.~A. and {Wander}, M. and {Wang}, L. and {Wang}, Y. and {Weaver}, J.~R. and {Weller}, J. and {Whalen}, D.~J. and {Wiesmann}, M. and {Wilde}, J. and {Williams}, O.~R. and {Winther}, H. -A. and {Wittje}, A. and {Wong}, J.~H.~W. and {Wright}, A.~H. and {Yankelevich}, V. and {Yeung}, H.~W. and {Youles}, S. and {Yung}, L.~Y.~A. and {Zacchei}, A. and {Zalesky}, L. and {Zamorani}, G. and {Zamorano Vitorelli}, A. and {Zanoni Marc}, M. and {Zennaro}, M. and {Zerbi}, F.~M. and {Zinchenko}, I.~A. and {Zoubian}, J. and {Zucca}, E. and {Zumalacarregui}, M.},
        title = "{Euclid. I. Overview of the Euclid mission}",
      journal = {arXiv e-prints},
         year = 2024,
        month = may,
          eid = {arXiv:2405.13491},
        pages = {arXiv:2405.13491},
          doi = {10.48550/arXiv.2405.13491},
archivePrefix = {arXiv},
       eprint = {2405.13491},
 primaryClass = {astro-ph.CO},
       adsurl = {https://ui.adsabs.harvard.edu/abs/2024arXiv240513491E}
}

@ARTICLE{EuclidNISP,
       author = {{Euclid Collaboration} and {Jahnke}, K. and {Gillard}, W. and {Schirmer}, M. and {Ealet}, A. and {Maciaszek}, T. and {Prieto}, E. and {Barbier}, R. and {Bonoli}, C. and {Corcione}, L. and {Dusini}, S. and {Grupp}, F. and {Hormuth}, F. and {Ligori}, S. and {Martin}, L. and {Morgante}, G. and {Padilla}, C. and {Toledo-Moreo}, R. and {Trifoglio}, M. and {Valenziano}, L. and {Bender}, R. and {Castander}, F.~J. and {Garilli}, B. and {Lilje}, P.~B. and {Rix}, H. -W. and {Auricchio}, N. and {Balestra}, A. and {Barriere}, J. -C. and {Battaglia}, P. and {Berthe}, M. and {Bodendorf}, C. and {Boenke}, T. and {Bon}, W. and {Bonnefoi}, A. and {Caillat}, A. and {Capobianco}, V. and {Carle}, M. and {Casas}, R. and {Cho}, H. and {Costille}, A. and {Ducret}, F. and {Ferriol}, S. and {Franceschi}, E. and {Gimenez}, J. -L. and {Holmes}, W. and {Hornstrup}, A. and {Jhabvala}, M. and {Kohley}, R. and {Kubik}, B. and {Laureijs}, R. and {Le Mignant}, D. and {Lloro}, I. and {Medinaceli}, E. and {Mellier}, Y. and {Polenta}, G. and {Racca}, G.~D. and {Renzi}, A. and {Salvignol}, J. -C. and {Secroun}, A. and {Seidel}, G. and {Seiffert}, M. and {Sirignano}, C. and {Sirri}, G. and {Strada}, P. and {Smadja}, G. and {Stanco}, L. and {Wachter}, S. and {Anselmi}, S. and {Borsato}, E. and {Caillat}, L. and {Cogato}, F. and {Colodro-Conde}, C. and {Crouzet}, P. -E. and {Conforti}, V. and {D'Alessandro}, M. and {Copin}, Y. and {Cuillandre}, J. -C. and {Davies}, J.~E. and {Davini}, S. and {Derosa}, A. and {Diaz}, J.~J. and {Di Domizio}, S. and {Di Ferdinando}, D. and {Farinelli}, R. and {Ferrari}, A.~G. and {Fornari}, F. and {Gabarra}, L. and {Gutierrez}, C.~M. and {Giacomini}, F. and {Lagier}, P. and {Gianotti}, F. and {Krause}, O. and {Madrid}, F. and {Laudisio}, F. and {Macias-Perez}, J. and {Naletto}, G. and {Niclas}, M. and {Marpaud}, J. and {Mauri}, N. and {da Silva}, R. and {Passalacqua}, F. and {Paterson}, K. and {Patrizii}, L. and {Risso}, I. and {Solheim}, B.~G.~B. and {Scodeggio}, M. and {Stassi}, P. and {Steinwagner}, J. and {Tenti}, M. and {Testera}, G. and {Travaglini}, R. and {Tosi}, S. and {Troja}, A. and {Tubio}, O. and {Valieri}, C. and {Vescovi}, C. and {Ventura}, S. and {Aghanim}, N. and {Altieri}, B. and {Amara}, A. and {Amiaux}, J. and {Andreon}, S. and {Aussel}, H. and {Baldi}, M. and {Bardelli}, S. and {Basset}, A. and {Bonchi}, A. and {Bonino}, D. and {Branchini}, E. and {Brescia}, M. and {Brinchmann}, J. and {Camera}, S. and {Carbone}, C. and {Cardone}, V.~F. and {Carretero}, J. and {Casas}, S. and {Castellano}, M. and {Cavuoti}, S. and {Chabaud}, P. -Y. and {Cimatti}, A. and {Congedo}, G. and {Conselice}, C.~J. and {Conversi}, L. and {Courbin}, F. and {Courtois}, H.~M. and {Cropper}, M. and {Cuby}, J. -G. and {Da Silva}, A. and {Degaudenzi}, H. and {Di Giorgio}, A.~M. and {Dinis}, J. and {Douspis}, M. and {Dubath}, F. and {Duncan}, C.~A.~J. and {Dupac}, X. and {Fabricius}, M. and {Farina}, M. and {Farrens}, S. and {Faustini}, F. and {Fosalba}, P. and {Fotopoulou}, S. and {Fourmanoit}, N. and {Frailis}, M. and {Franzetti}, P. and {Galeotta}, S. and {Gillis}, B. and {Giocoli}, C. and {G{\'o}mez-Alvarez}, P. and {Granett}, B.~R. and {Grazian}, A. and {Guzzo}, L. and {Hailey}, M. and {Haugan}, S.~V.~H. and {Hoar}, J. and {Hoekstra}, H. and {Hook}, I. and {Hudelot}, P. and {Joachimi}, B. and {Keih{\"a}nen}, E. and {Kermiche}, S. and {Kiessling}, A. and {Kilbinger}, M. and {Kitching}, T. and {K{\"u}mmel}, M. and {Kunz}, M. and {Kurki-Suonio}, H. and {Lahav}, O. and {Lindholm}, V. and {Alvarez}, J. Lorenzo and {Maino}, D. and {Maiorano}, E. and {Mansutti}, O. and {Marggraf}, O. and {Markovic}, K. and {Martignac}, J. and {Martinet}, N. and {Marulli}, F. and {Massey}, R. and {Masters}, D.~C. and {Maurogordato}, S. and {McCracken}, H.~J. and {Mei}, S. and {Melchior}, M. and {Meneghetti}, M. and {Merlin}, E. and {Meylan}, G. and {Mohr}, J.~J. and {Moresco}, M. and {Moscardini}, L. and {Nakajima}, R. and {Nichol}, R.~C. and {Niemi}, S. -M. and {Nutma}, T. and {Paech}, K. and {Paltani}, S. and {Pasian}, F. and {Peacock}, J.~A. and {Pedersen}, K. and {Percival}, W.~J. and {Pettorino}, V. and {Pires}, S. and {Poncet}, M. and {Popa}, L.~A. and {Pozzetti}, L. and {Raison}, F. and {Rebolo}, R. and {Refregier}, A. and {Rhodes}, J. and {Riccio}, G. and {Romelli}, E. and {Roncarelli}, M. and {Rosset}, C. and {Rossetti}, E. and {Rottgering}, H.~J.~A. and {Saglia}, R. and {Sapone}, D. and {Sauvage}, M. and {Scaramella}, R. and {Schneider}, P. and {Schrabback}, T. and {Serrano}, S. and {Tallada-Cresp{\'\i}}, P. and {Tavagnacco}, D. and {Taylor}, A.~N. and {Teplitz}, H.~I. and {Tereno}, I. and {Torradeflot}, F. and {Tutusaus}, I. and {Vassallo}, T. and {Verdoes Kleijn}, G. and {Veropalumbo}, A. and {Vibert}, D. and {Wang}, Y. and {Weller}, J. and {Zacchei}, A. and {Zamorani}, G. and {Zerbi}, F.~M. and {Zoubian}, J. and {Zucca}, E. and {Appleton}, P.~N. and {Baccigalupi}, C. and {Biviano}, A. and {Bolzonella}, M. and {Boucaud}, A. and {Bozzo}, E. and {Burigana}, C. and {Calabrese}, M. and {Casenove}, P. and {Crocce}, M. and {De Lucia}, G. and {Escartin Vigo}, J.~A. and {Fabbian}, G. and {Finelli}, F. and {George}, K. and {Gracia-Carpio}, J. and {Ili{\'c}}, S. and {Liebing}, P. and {Liu}, C. and {Mainetti}, G. and {Marcin}, S. and {Martinelli}, M. and {Morris}, P.~W. and {Neissner}, C. and {Pezzotta}, A. and {P{\"o}ntinen}, M. and {Porciani}, C. and {Sakr}, Z. and {Scottez}, V. and {Sefusatti}, E. and {Viel}, M. and {Wiesmann}, M. and {Akrami}, Y. and {Allevato}, V. and {Aubourg}, E. and {Ballardini}, M. and {Bertacca}, D. and {Bethermin}, M. and {Blanchard}, A. and {Blot}, L. and {Borgani}, S. and {Borlaff}, A.~S. and {Bruton}, S. and {Cabanac}, R. and {Calabro}, A. and {Calderone}, G. and {Canas-Herrera}, G. and {Cappi}, A. and {Carvalho}, C.~S. and {Castignani}, G. and {Castro}, T. and {Chambers}, K.~C. and {Charles}, Y. and {Chary}, R. and {Colbert}, J. and {Contarini}, S. and {Contini}, T. and {Cooray}, A.~R. and {Costanzi}, M. and {Cucciati}, O. and {De Caro}, B. and {de la Torre}, S. and {Desprez}, G. and {D{\'\i}az-S{\'a}nchez}, A. and {Dole}, H. and {Escoffier}, S. and {Ferreira}, P.~G. and {Ferrero}, I. and {Finoguenov}, A. and {Fontana}, A. and {Ganga}, K. and {Garc{\'\i}a-Bellido}, J. and {Gautard}, V. and {Gaztanaga}, E. and {Gozaliasl}, G. and {Gregorio}, A. and {Hall}, A. and {Hartley}, W.~G. and {Hemmati}, S. and {Hildebrandt}, H. and {Hjorth}, J. and {Hosseini}, S. and {Huertas-Company}, M. and {Ilbert}, O. and {Jacobson}, J. and {Joudaki}, S. and {Kajava}, J.~J.~E. and {Kansal}, V. and {Karagiannis}, D. and {Kirkpatrick}, C.~C. and {Lacasa}, F. and {Le Brun}, V. and {Le Graet}, J. and {Legrand}, L. and {Libet}, G. and {Liu}, S.~J. and {Loureiro}, A. and {Magliocchetti}, M. and {Mancini}, C. and {Mannucci}, F. and {Maoli}, R. and {Martins}, C.~J.~A.~P. and {Matthew}, S. and {Maurin}, L. and {McPartland}, C.~J.~R. and {Metcalf}, R.~B. and {Migliaccio}, M. and {Miluzio}, M. and {Monaco}, P. and {Moretti}, C. and {Nadathur}, S. and {Nicastro}, L. and {Walton}, Nicholas A. and {Odier}, J. and {Oguri}, M. and {Popa}, V. and {Potter}, D. and {Pourtsidou}, A. and {Rocci}, P. -F. and {Rollins}, R.~P. and {Rusholme}, B. and {Sahl{\'e}n}, M. and {S{\'a}nchez}, A.~G. and {Scarlata}, C. and {Schaye}, J. and {Schewtschenko}, J.~A. and {Schneider}, A. and {Schultheis}, M. and {Sereno}, M. and {Shankar}, F. and {Shulevski}, A. and {Sikkema}, G. and {Silvestri}, A. and {Simon}, P. and {Spurio Mancini}, A. and {Stadel}, J. and {Stanford}, S.~A. and {Tanidis}, K. and {Tao}, C. and {Tessore}, N. and {Teyssier}, R. and {Toft}, S. and {Tucci}, M. and {Valiviita}, J. and {Vergani}, D. and {Vernizzi}, F. and {Verza}, G. and {Vielzeuf}, P. and {Weaver}, J.~R. and {Zalesky}, L. and {Zinchenko}, I.~A. and {Archidiacono}, M. and {Atrio-Barandela}, F. and {Bennett}, C.~L. and {Bouvard}, T. and {Caro}, F. and {Conseil}, S. and {Dimauro}, P. and {Duc}, P. -A. and {Fang}, Y. and {Ferguson}, A.~M.~N. and {Gasparetto}, T. and {Kova\{{\v{c}}\}i{\'c}}, I. and {Kruk}, S. and {Le Brun}, A.~M.~C. and {Liaudat}, T.~I. and {Montoro}, A. and {Mora}, A. and {Murray}, C. and {Pagano}, L. and {Paoletti}, D. and {Radovich}, M. and {Sarpa}, E. and {Tommasi}, E. and {Viitanen}, A. and {Lesgourgues}, J. and {Levi}, M.~E. and {Mart{\'\i}n-Fleitas}, J.},
        title = "{Euclid. III. The NISP Instrument}",
      journal = {arXiv e-prints},
         year = 2024,
        month = may,
          eid = {arXiv:2405.13493},
        pages = {arXiv:2405.13493},
          doi = {10.48550/arXiv.2405.13493},
archivePrefix = {arXiv},
       eprint = {2405.13493},
 primaryClass = {astro-ph.IM},
       adsurl = {https://ui.adsabs.harvard.edu/abs/2024arXiv240513493E}
}

@article{Pfeifer_2022,
   title={COWS: a filament finder for Hessian cosmic web identifiers},
   volume={514},
   ISSN={1365-2966},
   url={http://dx.doi.org/10.1093/mnras/stac1382},
   DOI={10.1093/mnras/stac1382},
   number={1},
   journal={Monthly Notices of the Royal Astronomical Society},
   publisher={Oxford University Press (OUP)},
   author={Pfeifer, Simon and Libeskind, Noam I and Hoffman, Yehuda and Hellwing, Wojciech A and Bilicki, Maciej and Naidoo, Krishna},
   year={2022},
   month=may, pages={470–479} }

@article{Hahn_2007,
    author = {Hahn, Oliver and Porciani, Cristiano and Carollo, C. Marcella and Dekel, Avishai},
    title = {Properties of dark matter haloes in clusters, filaments, sheets and voids},
    journal = {Monthly Notices of the Royal Astronomical Society},
    volume = {375},
    number = {2},
    pages = {489-499},
    year = {2007},
    month = {01},
    issn = {0035-8711},
    doi = {10.1111/j.1365-2966.2006.11318.x},
    url = {https://doi.org/10.1111/j.1365-2966.2006.11318.x},
    eprint = {https://academic.oup.com/mnras/article-pdf/375/2/489/4244383/mnras0375-0489.pdf},
}

@ARTICLE{Shen_2006,
       author = {{Shen}, Jiajian and {Abel}, Tom and {Mo}, H.~J. and {Sheth}, Ravi K.},
        title = "{An Excursion Set Model of the Cosmic Web: The Abundance of Sheets, Filaments, and Halos}",
      journal = {\apj},
         year = 2006,
        month = jul,
       volume = {645},
       number = {2},
        pages = {783-791},
          doi = {10.1086/504513},
archivePrefix = {arXiv},
       eprint = {astro-ph/0511365},
 primaryClass = {astro-ph},
       adsurl = {https://ui.adsabs.harvard.edu/abs/2006ApJ...645..783S}
}

@ARTICLE{doroshkevich:1970,
       author = {{Doroshkevich}, A.~G.},
        title = "{Spatial structure of perturbations and origin of galactic rotation in fluctuation theory}",
      journal = {Astrophysics},
         year = 1970,
        month = oct,
       volume = {6},
       number = {4},
        pages = {320-330},
          doi = {10.1007/BF01001625},
       adsurl = {https://ui.adsabs.harvard.edu/abs/1970Ap......6..320D}
}

@ARTICLE{bbks,
       author = {{Bardeen}, J.~M. and {Bond}, J.~R. and {Kaiser}, N. and {Szalay}, A.~S.},
        title = "{The Statistics of Peaks of Gaussian Random Fields}",
      journal = {\apj},
         year = 1986,
        month = may,
       volume = {304},
        pages = {15},
          doi = {10.1086/164143},
       adsurl = {https://ui.adsabs.harvard.edu/abs/1986ApJ...304...15B}
}

@ARTICLE{jing/suto:2002,
       author = {{Jing}, Y.~P. and {Suto}, Yasushi},
        title = "{Triaxial Modeling of Halo Density Profiles with High-Resolution N-Body Simulations}",
      journal = {\apj},
         year = 2002,
        month = aug,
       volume = {574},
       number = {2},
        pages = {538-553},
          doi = {10.1086/341065},
archivePrefix = {arXiv},
       eprint = {astro-ph/0202064},
 primaryClass = {astro-ph},
       adsurl = {https://ui.adsabs.harvard.edu/abs/2002ApJ...574..538J}
}

@ARTICLE{hoffman:1986,
       author = {{Hoffman}, Y.},
        title = "{The Dynamics of Superclusters: The Effect of Shear}",
      journal = {\apj},
         year = 1986,
        month = sep,
       volume = {308},
        pages = {493},
          doi = {10.1086/164520},
       adsurl = {https://ui.adsabs.harvard.edu/abs/1986ApJ...308..493H}
}

@ARTICLE{peebles:1990,
       author = {{Peebles}, P.~J.~E.},
        title = "{The Gravitational Instability Picture and the Formation of the Local Group}",
      journal = {\apj},
         year = 1990,
        month = oct,
       volume = {362},
        pages = {1},
          doi = {10.1086/169237},
       adsurl = {https://ui.adsabs.harvard.edu/abs/1990ApJ...362....1P}
}

@ARTICLE{dubinski:1992,
       author = {{Dubinski}, John},
        title = "{Cosmological Tidal Shear}",
      journal = {\apj},
         year = 1992,
        month = dec,
       volume = {401},
        pages = {441},
          doi = {10.1086/172076},
       adsurl = {https://ui.adsabs.harvard.edu/abs/1992ApJ...401..441D}
}

@ARTICLE{bertschinger/jain:1994,
       author = {{Bertschinger}, Edmund and {Jain}, Bhuvnesh},
        title = "{Gravitational Instability of Cold Matter}",
      journal = {\apj},
         year = 1994,
        month = aug,
       volume = {431},
        pages = {486},
          doi = {10.1086/174501},
archivePrefix = {arXiv},
       eprint = {astro-ph/9307033},
 primaryClass = {astro-ph},
       adsurl = {https://ui.adsabs.harvard.edu/abs/1994ApJ...431..486B}
}

@ARTICLE{bond/kofman/pogosyan:1996,
       author = {{Bond}, J. Richard and {Kofman}, Lev and {Pogosyan}, Dmitry},
        title = "{How filaments of galaxies are woven into the cosmic web}",
      journal = {\nat},
         year = 1996,
        month = apr,
       volume = {380},
       number = {6575},
        pages = {603-606},
          doi = {10.1038/380603a0},
archivePrefix = {arXiv},
       eprint = {astro-ph/9512141},
 primaryClass = {astro-ph},
       adsurl = {https://ui.adsabs.harvard.edu/abs/1996Natur.380..603B}
}

@ARTICLE{sheth/mo/tormen:2001,
       author = {{Sheth}, Ravi K. and {Mo}, H.~J. and {Tormen}, Giuseppe},
        title = "{Ellipsoidal collapse and an improved model for the number and spatial distribution of dark matter haloes}",
      journal = {\mnras},
         year = 2001,
        month = may,
       volume = {323},
       number = {1},
        pages = {1-12},
          doi = {10.1046/j.1365-8711.2001.04006.x},
archivePrefix = {arXiv},
       eprint = {astro-ph/9907024},
 primaryClass = {astro-ph},
       adsurl = {https://ui.adsabs.harvard.edu/abs/2001MNRAS.323....1S}
}

@ARTICLE{desjacques:2008,
       author = {{Desjacques}, Vincent},
        title = "{Environmental dependence in the ellipsoidal collapse model}",
      journal = {\mnras},
         year = 2008,
        month = aug,
       volume = {388},
       number = {2},
        pages = {638-658},
          doi = {10.1111/j.1365-2966.2008.13420.x},
archivePrefix = {arXiv},
       eprint = {0707.4670},
 primaryClass = {astro-ph},
       adsurl = {https://ui.adsabs.harvard.edu/abs/2008MNRAS.388..638D}
}

@ARTICLE{paranjape/etal:2018,
       author = {{Paranjape}, Aseem and {Hahn}, Oliver and {Sheth}, Ravi K.},
        title = "{Halo assembly bias and the tidal anisotropy of the local halo environment}",
      journal = {\mnras},
         year = 2018,
        month = may,
       volume = {476},
       number = {3},
        pages = {3631-3647},
          doi = {10.1093/mnras/sty496},
archivePrefix = {arXiv},
       eprint = {1706.09906},
 primaryClass = {astro-ph.CO},
       adsurl = {https://ui.adsabs.harvard.edu/abs/2018MNRAS.476.3631P}
}

@ARTICLE{paranjape:2021,
       author = {{Paranjape}, Aseem},
        title = "{Analytical halo models of cosmic tidal fields}",
      journal = {\mnras},
         year = 2021,
        month = apr,
       volume = {502},
       number = {4},
        pages = {5210-5226},
          doi = {10.1093/mnras/stab359},
archivePrefix = {arXiv},
       eprint = {2006.13954},
 primaryClass = {astro-ph.CO},
       adsurl = {https://ui.adsabs.harvard.edu/abs/2021MNRAS.502.5210P}
}

@ARTICLE{favole/etal:2022,
       author = {{Favole}, Ginevra and {Montero-Dorta}, Antonio D. and {Artale}, M. Celeste and {Contreras}, Sergio and {Zehavi}, Idit and {Xu}, Xiaoju},
        title = "{Subhalo abundance matching through the lens of a hydrodynamical simulation}",
      journal = {\mnras},
         year = 2022,
        month = jan,
       volume = {509},
       number = {2},
        pages = {1614-1625},
          doi = {10.1093/mnras/stab3006},
archivePrefix = {arXiv},
       eprint = {2101.10733},
 primaryClass = {astro-ph.GA},
       adsurl = {https://ui.adsabs.harvard.edu/abs/2022MNRAS.509.1614F}
}

@ARTICLE{osato/okumura:2023,
       author = {{Osato}, Ken and {Okumura}, Teppei},
        title = "{Clustering of emission line galaxies with IllustrisTNG - I. Fundamental properties and halo occupation distribution}",
      journal = {\mnras},
         year = 2023,
        month = feb,
       volume = {519},
       number = {2},
        pages = {1771-1791},
          doi = {10.1093/mnras/stac3582},
archivePrefix = {arXiv},
       eprint = {2206.08678},
 primaryClass = {astro-ph.GA},
       adsurl = {https://ui.adsabs.harvard.edu/abs/2023MNRAS.519.1771O}
}

@ARTICLE{foreroromero/etal:2009,
       author = {{Forero-Romero}, J.~E. and {Hoffman}, Y. and {Gottl{\"o}ber}, S. and {Klypin}, A. and {Yepes}, G.},
        title = "{A dynamical classification of the cosmic web}",
      journal = {\mnras},
         year = 2009,
        month = jul,
       volume = {396},
       number = {3},
        pages = {1815-1824},
          doi = {10.1111/j.1365-2966.2009.14885.x},
archivePrefix = {arXiv},
       eprint = {0809.4135},
 primaryClass = {astro-ph},
       adsurl = {https://ui.adsabs.harvard.edu/abs/2009MNRAS.396.1815F}
}

@ARTICLE{martizzi/etal:2019,
       author = {{Martizzi}, Davide and {Vogelsberger}, Mark and {Artale}, Maria Celeste and {Haider}, Markus and {Torrey}, Paul and {Marinacci}, Federico and {Nelson}, Dylan and {Pillepich}, Annalisa and {Weinberger}, Rainer and {Hernquist}, Lars and {Naiman}, Jill and {Springel}, Volker},
        title = "{Baryons in the Cosmic Web of IllustrisTNG - I: gas in knots, filaments, sheets, and voids}",
      journal = {\mnras},
         year = 2019,
        month = jul,
       volume = {486},
       number = {3},
        pages = {3766-3787},
          doi = {10.1093/mnras/stz1106},
archivePrefix = {arXiv},
       eprint = {1810.01883},
 primaryClass = {astro-ph.CO},
       adsurl = {https://ui.adsabs.harvard.edu/abs/2019MNRAS.486.3766M}
}

@ARTICLE{okane/etal:2024,
       author = {{O'Kane}, Callum J. and {Kuchner}, Ulrike and {Gray}, Meghan E. and {Arag{\'o}n-Salamanca}, Alfonso},
        title = "{The effect of cosmic web filaments on galaxy evolution}",
      journal = {\mnras},
         year = 2024,
        month = nov,
       volume = {534},
       number = {3},
        pages = {1682-1699},
          doi = {10.1093/mnras/stae2142},
archivePrefix = {arXiv},
       eprint = {2409.09028},
 primaryClass = {astro-ph.GA},
       adsurl = {https://ui.adsabs.harvard.edu/abs/2024MNRAS.534.1682O}
}

@ARTICLE{rapoport/etal:2025,
       author = {{Rapoport}, Ivan and {Desjacques}, Vincent and {Parimbelli}, Gabriele and {Behar}, Ehud and {Crocce}, Martin},
        title = "{Modeling spatially-resolved galactic H$\alpha$ emission for galaxy clustering}",
      journal = {arXiv e-prints},
         year = 2025,
        month = feb,
          eid = {arXiv:2502.08778},
        pages = {arXiv:2502.08778},
          doi = {10.48550/arXiv.2502.08778},
archivePrefix = {arXiv},
       eprint = {2502.08778},
 primaryClass = {astro-ph.GA},
       adsurl = {https://ui.adsabs.harvard.edu/abs/2025arXiv250208778R}
}

@ARTICLE{abbas/sheth:2007,
       author = {{Abbas}, Ummi and {Sheth}, Ravi K.},
        title = "{Strong clustering of underdense regions and the environmental dependence of clustering from Gaussian initial conditions}",
      journal = {\mnras},
         year = 2007,
        month = jun,
       volume = {378},
       number = {2},
        pages = {641-648},
          doi = {10.1111/j.1365-2966.2007.11806.x},
archivePrefix = {arXiv},
       eprint = {astro-ph/0703391},
 primaryClass = {astro-ph},
       adsurl = {https://ui.adsabs.harvard.edu/abs/2007MNRAS.378..641A}
}

@ARTICLE{shi/sheth:2018,
       author = {{Shi}, Jingjing and {Sheth}, Ravi K.},
        title = "{Dependence of halo bias on mass and environment}",
      journal = {\mnras},
         year = 2018,
        month = jan,
       volume = {473},
       number = {2},
        pages = {2486-2492},
          doi = {10.1093/mnras/stx2277},
archivePrefix = {arXiv},
       eprint = {1707.04096},
 primaryClass = {astro-ph.CO},
       adsurl = {https://ui.adsabs.harvard.edu/abs/2018MNRAS.473.2486S}
}

@ARTICLE{repp/szapudi:2022,
       author = {{Repp}, Andrew and {Szapudi}, Istv{\'a}n},
        title = "{Indicator power spectra: surgical excision of non-linearities and covariance matrices for counts in cells}",
      journal = {\mnras},
         year = 2022,
        month = jan,
       volume = {509},
       number = {1},
        pages = {586-594},
          doi = {10.1093/mnras/stab3031},
archivePrefix = {arXiv},
       eprint = {2108.01673},
 primaryClass = {astro-ph.CO},
       adsurl = {https://ui.adsabs.harvard.edu/abs/2022MNRAS.509..586R}
}

@ARTICLE{kaiser:1984,
       author = {{Kaiser}, N.},
        title = "{On the spatial correlations of Abell clusters.}",
      journal = {\apjl},
         year = 1984,
        month = sep,
       volume = {284},
        pages = {L9-L12},
          doi = {10.1086/184341},
       adsurl = {https://ui.adsabs.harvard.edu/abs/1984ApJ...284L...9K}
}

@ARTICLE{cole/kaiser:1989,
       author = {{Cole}, Shaun and {Kaiser}, Nick},
        title = "{Biased clustering in the cold dark matter cosmogony.}",
      journal = {\mnras},
         year = 1989,
        month = apr,
       volume = {237},
        pages = {1127-1146},
          doi = {10.1093/mnras/237.4.1127},
       adsurl = {https://ui.adsabs.harvard.edu/abs/1989MNRAS.237.1127C}
}

@ARTICLE{mo/white:1996,
       author = {{Mo}, H.~J. and {White}, S.~D.~M.},
        title = "{An analytic model for the spatial clustering of dark matter haloes}",
      journal = {\mnras},
         year = 1996,
        month = sep,
       volume = {282},
       number = {2},
        pages = {347-361},
          doi = {10.1093/mnras/282.2.347},
archivePrefix = {arXiv},
       eprint = {astro-ph/9512127},
 primaryClass = {astro-ph},
       adsurl = {https://ui.adsabs.harvard.edu/abs/1996MNRAS.282..347M}
}

@article{Tumlinson_2017,
   title={The Circumgalactic Medium},
   volume={55},
   ISSN={1545-4282},
   url={http://dx.doi.org/10.1146/annurev-astro-091916-055240},
   DOI={10.1146/annurev-astro-091916-055240},
   number={1},
   journal={Annual Review of Astronomy and Astrophysics},
   publisher={Annual Reviews},
   author={Tumlinson, Jason and Peeples, Molly S. and Werk, Jessica K.},
   year={2017},
   month=aug, pages={389–432} }

@article{Hopkins_2012,
   title={Stellar feedback and bulge formation in clumpy discs: Feedback and clump coalescence},
   volume={427},
   ISSN={1365-2966},
   url={http://dx.doi.org/10.1111/j.1365-2966.2012.21981.x},
   DOI={10.1111/j.1365-2966.2012.21981.x},
   number={2},
   journal={Monthly Notices of the Royal Astronomical Society},
   publisher={Oxford University Press (OUP)},
   author={Hopkins, Philip F. and Kereš, Dusan and Murray, Norman and Quataert, Eliot and Hernquist, Lars},
   year={2012},
   month=nov, pages={968–978} }

@ARTICLE{Kennicutt_1983,
       author = {{Kennicutt}, R.~C., Jr.},
        title = "{The rate of star formation in normal disk galaxies.}",
      journal = {\apj},
         year = 1983,
        month = sep,
       volume = {272},
        pages = {54-67},
          doi = {10.1086/161261},
       adsurl = {https://ui.adsabs.harvard.edu/abs/1983ApJ...272...54K}
}

@ARTICLE{Keres2005,
       author = {{Kere{\v{s}}}, Du{\v{s}}an and {Katz}, Neal and {Weinberg}, David H. and {Dav{\'e}}, Romeel},
        title = "{How do galaxies get their gas?}",
      journal = {\mnras},
         year = 2005,
        month = oct,
       volume = {363},
       number = {1},
        pages = {2-28},
          doi = {10.1111/j.1365-2966.2005.09451.x},
archivePrefix = {arXiv},
       eprint = {astro-ph/0407095},
 primaryClass = {astro-ph},
       adsurl = {https://ui.adsabs.harvard.edu/abs/2005MNRAS.363....2K}
}

@article{TNG_1,
   title={First results from the IllustrisTNG simulations: the galaxy colour bimodality},
   volume={475},
   ISSN={1365-2966},
   url={http://dx.doi.org/10.1093/mnras/stx3040},
   DOI={10.1093/mnras/stx3040},
   number={1},
   journal={Monthly Notices of the Royal Astronomical Society},
   publisher={Oxford University Press (OUP)},
   author={Nelson, Dylan and Pillepich, Annalisa and Springel, Volker and Weinberger, Rainer and Hernquist, Lars and Pakmor, Rüdiger and Genel, Shy and Torrey, Paul and Vogelsberger, Mark and Kauffmann, Guinevere and Marinacci, Federico and Naiman, Jill},
   year={2017},
   month=nov, pages={624–647} }

@article{TNG_2,
   title={First results from the IllustrisTNG simulations: radio haloes and magnetic fields},
   ISSN={1365-2966},
   url={http://dx.doi.org/10.1093/mnras/sty2206},
   DOI={10.1093/mnras/sty2206},
   journal={Monthly Notices of the Royal Astronomical Society},
   publisher={Oxford University Press (OUP)},
   author={Marinacci, Federico and Vogelsberger, Mark and Pakmor, Rüdiger and Torrey, Paul and Springel, Volker and Hernquist, Lars and Nelson, Dylan and Weinberger, Rainer and Pillepich, Annalisa and Naiman, Jill and Genel, Shy},
   year={2018},
   month=aug }

@article{TNG_3,
   title={First results from the IllustrisTNG simulations: a tale of two elements – chemical evolution of magnesium and europium},
   volume={477},
   ISSN={1365-2966},
   url={http://dx.doi.org/10.1093/mnras/sty618},
   DOI={10.1093/mnras/sty618},
   number={1},
   journal={Monthly Notices of the Royal Astronomical Society},
   publisher={Oxford University Press (OUP)},
   author={Naiman, Jill P and Pillepich, Annalisa and Springel, Volker and Ramirez-Ruiz, Enrico and Torrey, Paul and Vogelsberger, Mark and Pakmor, Rüdiger and Nelson, Dylan and Marinacci, Federico and Hernquist, Lars and Weinberger, Rainer and Genel, Shy},
   year={2018},
   month=mar, pages={1206–1224} }

@article{TNG_4,
   title={First results from the IllustrisTNG simulations: matter and galaxy clustering},
   volume={475},
   ISSN={1365-2966},
   url={http://dx.doi.org/10.1093/mnras/stx3304},
   DOI={10.1093/mnras/stx3304},
   number={1},
   journal={Monthly Notices of the Royal Astronomical Society},
   publisher={Oxford University Press (OUP)},
   author={Springel, Volker and Pakmor, Rüdiger and Pillepich, Annalisa and Weinberger, Rainer and Nelson, Dylan and Hernquist, Lars and Vogelsberger, Mark and Genel, Shy and Torrey, Paul and Marinacci, Federico and Naiman, Jill},
   year={2017},
   month=dec, pages={676–698} }

@article{TNG_5,
   title={First results from the IllustrisTNG simulations: the stellar mass content of groups and clusters of galaxies},
   volume={475},
   ISSN={1365-2966},
   url={http://dx.doi.org/10.1093/mnras/stx3112},
   DOI={10.1093/mnras/stx3112},
   number={1},
   journal={Monthly Notices of the Royal Astronomical Society},
   publisher={Oxford University Press (OUP)},
   author={Pillepich, Annalisa and Nelson, Dylan and Hernquist, Lars and Springel, Volker and Pakmor, Rüdiger and Torrey, Paul and Weinberger, Rainer and Genel, Shy and Naiman, Jill P and Marinacci, Federico and Vogelsberger, Mark},
   year={2017},
   month=dec, pages={648–675} }

@article{TNG50a,
    author = {Nelson, Dylan and Pillepich, Annalisa and Springel, Volker and Pakmor, Rüdiger and Weinberger, Rainer and Genel, Shy and Torrey, Paul and Vogelsberger, Mark and Marinacci, Federico and Hernquist, Lars},
    title = {First results from the TNG50 simulation: galactic outflows driven by supernovae and black hole feedback},
    journal = {Monthly Notices of the Royal Astronomical Society},
    volume = {490},
    number = {3},
    pages = {3234-3261},
    year = {2019},
    month = {08},
    issn = {0035-8711},
    doi = {10.1093/mnras/stz2306},
    url = {https://doi.org/10.1093/mnras/stz2306},
    eprint = {https://academic.oup.com/mnras/article-pdf/490/3/3234/30327914/stz2306.pdf},
}

@article{TNG50b,
    author = {Pillepich, Annalisa and Nelson, Dylan and Springel, Volker and Pakmor, Rüdiger and Torrey, Paul and Weinberger, Rainer and Vogelsberger, Mark and Marinacci, Federico and Genel, Shy and van der Wel, Arjen and Hernquist, Lars},
    title = {First results from the TNG50 simulation: the evolution of stellar and gaseous discs across cosmic time},
    journal = {Monthly Notices of the Royal Astronomical Society},
    volume = {490},
    number = {3},
    pages = {3196-3233},
    year = {2019},
    month = {09},
    issn = {0035-8711},
    doi = {10.1093/mnras/stz2338},
    url = {https://doi.org/10.1093/mnras/stz2338},
    eprint = {https://academic.oup.com/mnras/article-pdf/490/3/3196/30327933/stz2338.pdf},
}

@misc{TNG_DR,
      title={The IllustrisTNG Simulations: Public Data Release}, 
      author={Dylan Nelson and Volker Springel and Annalisa Pillepich and Vicente Rodriguez-Gomez and Paul Torrey and Shy Genel and Mark Vogelsberger and Ruediger Pakmor and Federico Marinacci and Rainer Weinberger and Luke Kelley and Mark Lovell and Benedikt Diemer and Lars Hernquist},
      year={2021},
      eprint={1812.05609},
      archivePrefix={arXiv},
      primaryClass={astro-ph.GA},
      url={https://arxiv.org/abs/1812.05609}, 
}

@book{Draine2011,
  author    = {Draine, Bruce T.},
  title     = {Physics of the Interstellar and Intergalactic Medium},
  publisher = {Princeton University Press},
  year      = {2011},
  address   = {Princeton, NJ},
  isbn      = {978-0-691-12214-4}
}

@article{Darvish_2014,
   title={COSMIC WEB AND STAR FORMATION ACTIVITY IN GALAXIES ATz∼ 1},
   volume={796},
   ISSN={1538-4357},
   url={http://dx.doi.org/10.1088/0004-637X/796/1/51},
   DOI={10.1088/0004-637x/796/1/51},
   number={1},
   journal={The Astrophysical Journal},
   publisher={American Astronomical Society},
   author={Darvish, B. and Sobral, D. and Mobasher, B. and Scoville, N. Z. and Best, P. and Sales, L. V. and Smail, I.},
   year={2014},
   month=nov, pages={51} }

@ARTICLE{Yang2009,
       author = {{Yang}, Xiaohu and {Mo}, H.~J. and {van den Bosch}, Frank C.},
        title = "{Galaxy Groups in the SDSS DR4. III. The Luminosity and Stellar Mass Functions}",
      journal = {\apj},
         year = 2009,
        month = apr,
       volume = {695},
       number = {2},
        pages = {900-916},
          doi = {10.1088/0004-637X/695/2/900},
archivePrefix = {arXiv},
       eprint = {0808.0539},
 primaryClass = {astro-ph},
       adsurl = {https://ui.adsabs.harvard.edu/abs/2009ApJ...695..900Y}
}

@article{Baldry2006,
    author = {Baldry, I. K. and Balogh, M. L. and Bower, R. G. and Glazebrook, K. and Nichol, R. C. and Bamford, S. P. and Budavari, T.},
    title = {Galaxy bimodality versus stellar mass and environment},
    journal = {Monthly Notices of the Royal Astronomical Society},
    volume = {373},
    number = {2},
    pages = {469-483},
    year = {2006},
    month = {10},
    issn = {0035-8711},
    doi = {10.1111/j.1365-2966.2006.11081.x},
    url = {https://doi.org/10.1111/j.1365-2966.2006.11081.x},
    eprint = {https://academic.oup.com/mnras/article-pdf/373/2/469/4099171/mnras0373-0469.pdf},
}

\end{document}